\documentclass[lettersize,journal]{IEEEtran}
\usepackage{amsmath,amsfonts}
\usepackage{algorithmic}
\usepackage [ruled,linesnumbered]{algorithm2e}
\usepackage{array}
\usepackage[caption=false]{subfig}
\usepackage{textcomp}
\usepackage{stfloats}
\usepackage{url}
\usepackage{verbatim}
\usepackage{graphicx}
\usepackage{cite}
\usepackage{xcolor}
\usepackage{bm}
\usepackage{multirow}
\usepackage[most]{tcolorbox}
\usepackage[colorlinks,
            linkcolor=blue,
            anchorcolor=black,
            citecolor=blue,
            ]{hyperref}
\usepackage{hyperref}
\usepackage{threeparttable}
\usepackage{array}
\usepackage{makecell}

\newcommand{\methodname}[1]{\textsc{ReIC}}
\newenvironment{RQbox}[2]{
\begin{center}
    \begin{tcolorbox}[
        colback=gray!30,
    	colframe=black,					
    	width=9cm,							
    	arc=1mm, auto outer arc,
    	boxrule=1pt]
    	\textbf{\textit{Conclusion for #1}}: #2
     
    \end{tcolorbox}
\end{center}}
{}

\begin{document}
\title{Metamorphic Testing of Image Captioning Systems via Image-Level Reduction}

\author{
\IEEEauthorblockN{
Xiaoyuan Xie\IEEEauthorrefmark{2}\IEEEauthorrefmark{1}\thanks{\IEEEauthorrefmark{1}Xiaoyuan Xie and Songqiang Chen are the co-corresponding authors.},
Xingpeng Li\IEEEauthorrefmark{2},
Songqiang Chen\IEEEauthorrefmark{3}\IEEEauthorrefmark{1}
}

\IEEEauthorblockA{\IEEEauthorrefmark{2}School of Computer Science, Wuhan University, China}

\IEEEauthorblockA{\IEEEauthorrefmark{3}Department of Computer Science and Engineering, The Hong Kong University of Science and Technology, China}

\IEEEauthorblockA{xxie@whu.edu.cn, lixingpeng@whu.edu.cn, i9s.chen@connect.ust.hk}
}

\maketitle
\begin{abstract}
The Image Captioning (IC) technique is widely used to describe images in natural language. However, even state-of-the-art IC systems can still produce incorrect captions and lead to misunderstandings. Recently, some IC system testing methods have been proposed. However, these methods still rely on pre-annotated information and hence cannot really alleviate the difficulty in identifying the test oracle. Furthermore, their methods artificially manipulate objects, which may generate unreal images as test cases and thus lead to less meaningful testing results. Thirdly, existing methods have various requirements on the eligibility of source test cases, and hence cannot fully utilize the given images to perform testing.
To tackle these issues, in this paper, we propose \methodname{} to perform metamorphic testing for the IC systems with some image-level reduction transformations like image cropping and stretching. Instead of relying on the pre-annotated information, \methodname{} uses a localization method to align objects in the caption with corresponding objects in the image, and checks whether each object is correctly described or deleted in the caption after transformation. With the image-level reduction transformations, \methodname{} does not artificially manipulate any objects and hence can avoid generating unreal follow-up images. Additionally, it eliminates the requirement on the eligibility of source test cases during the metamorphic transformation process, as well as decreases the ambiguity and boosts the diversity among the follow-up test cases, which consequently enables testing to be performed on any test image and reveals more distinct valid violations. 
We employ \methodname{} to test five popular IC systems. The results demonstrate that \methodname{} can sufficiently leverage the provided test images to generate follow-up cases of good realism, and effectively detect a great number of distinct violations, without the need for any pre-annotated information.

\end{abstract}


\nocite{DATARELEASE}
\nocite{RobustICCode}
\nocite{azure}
\section{Introduction}
\label{sec:introduction}

\IEEEPARstart{W}{ith} the development of computer vision and natural language techniques, the Image Captioning (IC) technique has been proposed to describe the salient objects of an image in natural language~\cite{icsurvey, showandtell, Oscar, VinVL, XLAN, OFA}. IC systems\footnote{Following Yu et al.~\cite{MetaIC_Pinjia}, we consider IC models and commercial IC services as ``IC systems'' in this paper.} have been widely used in modern society for many aims, such as helping visually impaired people to understand the real world~\cite{icsurvey, blindmen}. However, similar to other AI-driven systems, the IC systems also manifest many defects that can lead to erroneous output captions. Such incorrect captions could mislead users and result in their inappropriate actions, e.g., users may comment on social media platforms based on the wrong understanding~\cite{ICquality, icquality1, icquality2}. To reveal the potential issues in the IC systems, it is essential to conduct sufficient testing with effective methods.

To automatically test the IC systems, Yu et al. proposed the first automated IC system testing method, MetaIC~\cite{MetaIC_Pinjia}. MetaIC inserts an object into the original images to generate new images and checks if the IC Software Under Test (SUT) correctly depicts all the objects in the source caption plus the object inserted in the follow-up test case. Later, Yu et al. further proposed ROME~\cite{ROME_Pinjia} that generated follow-up test cases by first removing objects from the original images and then inpainting the images, trying to ameliorate the weakness of MetaIC in generating many invalid and unrealistic follow-up test cases.

However, the existing methods mentioned above still suffer from three major limitations. {Firstly}, as a method that aims to resolve the \textit{oracle problem}~\cite{oraclesurvey} in IC testing activity, neither MetaIC nor ROME completely cuts off their dependency on the annotations. These approaches still rely on some annotated information about the images, including the categories of objects in the image with their position and corresponding masks~\cite{RobustICCode}. As a result, the cost of extensive human efforts on annotations will not be really saved in these approaches as compared with traditional reference-based testing approaches.
Secondly, even though ROME claims that it can generate more realistic images than MetaIC because it melts objects from the image instead of inserting objects into the image, it still suffers from the problem of generating unreal follow-up images, such as partially melting an object from its surroundings.
{Thirdly}, existing methods have strict requirements on the eligibility of source test cases for their transformations. For example, MetaIC selects the images whose existing objects are not too large to guarantee sufficient space for inserting new objects as the eligible test cases. As a result, many available images cannot be leveraged to conduct testing.

In this work, we propose a new testing method, \methodname{}, which adopts image-level \textsc{Re}duction transformations for the \textsc{IC} systems.
\methodname{} conducts Metamorphic Testing (MT)~\cite{MTConcept} based on a series of transformations that do not suffer from the aforementioned limitations. 
Specifically, to dispense the first limitation about the reliance on annotated information, \methodname{} establishes the correspondence between the words in the caption and the objects in the test image. 
It uses an Object Detection (OD) model together with a newly designed occlusion-based localization method that utilizes the captions of occluded images generated by the IC system to estimate the position of all the objects described in the source caption within the image. 
In this way, \methodname{} eliminates the reliance on the annotations of test images, thereby addressing \textit{the first limitation}.

Furthermore, unlike existing works that try to manipulate particular objects in an image to generate new test cases, we adopt the reduction-based transformation \textbf{on the whole image}, such as cropping, stretching, and rotation\footnote{During the stretching transformation, the portion of the image that exceeds the image boundary is removed. Similarly, during the rotation transformation, to avoid unrealistic black board, the black border and the surrounding area are also removed. Further details can be found in Section~\ref{subsec:MRs}.}. Since \methodname{} \textbf{does not artificially manipulate any objects}, it will preserve the original naturalness of the image. \methodname{} will not bring in any potential unrealistic factors, consequently addressing \textit{the second limitation.} At the same time, by using direct reduction-based transformations on the entire image, \methodname{} \textbf{has no requirement on the eligibility of} source test cases. As a result, it is able to perform transformations without any restriction during the process and hence can address \textit{the last limitation.}

As a reminder, in \methodname{}, the expected relation between source and follow-up outputs depends on whether each object in the source caption is reduced after the transformation. However, when applying reduction-based transformations to the images, some objects may be \textbf{partially} cut. It is hard to determine the expected model behavior for these ``ambiguous'' objects. To avoid such ambiguous objects, we further design a measure, namely, $Score_{ambiguity}$ to identify and exclude such ambiguous cases. Moreover, we notice that the number of reported violations cannot comprehensively demonstrate the capability of a testing method. In fact, multiple violations can be due to quite similar causes. Thus, we propose another measure, namely, $Score_{diversity}$ to essentially increase the number of revealed \textbf{distinct} violations.
We design a dynamic strategy to generate follow-up images for \textbf{higher diversity} and \textbf{lower ambiguity}, and hence lead to better fault detection.

We apply \methodname{} to test four typical IC models~\cite{showandtell, Oscar, VinVL, OFA} and a commercial IC service~\cite{azure}. Experimental results demonstrate that \methodname{} is very effective in revealing true violations using \textbf{unannotated images} for popular IC systems; while existing methods~\cite{MetaIC_Pinjia, ROME_Pinjia} strictly rely on pre-annotated information. Additionally, \methodname{} can fully utilize all provided images to construct follow-up test cases. In contrast, existing methods have a series of restrictions on selecting eligible source images. The test cases generated by \methodname{} are also found to be more realistic than those generated by existing approaches, indicating that the detected errors hold greater value.

In summary, this work makes the following contributions:

\begin{itemize}

    \item We introduce an automated method named \methodname{}, to test the IC systems. Our method does not refer to any pre-annotated information of test images for generating follow-up images and detecting violations.

    \item Our method designs a dynamic strategy and employs reduction-based transformation, which does not artificially manipulate any objects and can preserve the original naturalness of the image. It effectively avoids generating unreal follow-up test cases. Furthermore, our method eliminates the restrictions during the metamorphic transformation process, as well as decreases the ambiguity, and boosts the diversity among the follow-up test cases, which consequently enables testing to be performed on any test image, and reveals more distinct valid violations.
    
    \item We have designed a tool and implemented three reduction-based transformations (i.e., cropping, stretching, and rotation) to test the IC systems. We conduct comprehensive experiments on five IC systems using two datasets. The results demonstrate that \methodname{} can sufficiently leverage provided test images to generate follow-up cases of good realism, and effectively detect a great number of distinct violations, without the need for any pre-annotated information.
	
\end{itemize}

The artifact of this paper is available at~\cite{DATARELEASE}.

The rest of this paper goes as follows. Section~\ref{sec:Background} introduces the IC technique, MT technique and the latest methods for IC testing. Section~\ref{sec:Preliminaries} introduces the motivation and challenge of this work. Section~\ref{sec:Methodology} elaborates our methodology in detail. After that, Sections~\ref{sec:exp} and \ref{sec:Results} present the setup and the results of our evaluation experiments, respectively. Next, Section~\ref{sec:discuss} discusses the issue of realism, false positives, efficiency of \methodname{} and threats to validity, and Section~\ref{sec:RelatedWork} lists the related works. Finally, Section~\ref{sec:Conclusion} concludes this paper.

\section{Background}
\label{sec:Background}

\subsection{Image Captioning}
\label{subsec:icintro}
Image captioning is an image-to-text task that describes the salient content of an image in natural language. Typically, IC tasks involve two primary components, visual encoding and language modeling~\cite{icsurvey}. Firstly, images are encoded as one or multiple feature vectors, which serve as the input for the language model. Then, the language model generates a sequence of words or sub-words decoded according to a provided vocabulary. Currently, the approaches for visual encoding can be classified into the following types: (1) CNN-based features~\cite{showandtell}; (2) visual contents over grids or regions~\cite{visualGridEncoder, bottomup}; (3) graph-based methods adding visual relationships between visual regions~\cite{graphEncoder}; and (4) Transformer-based methods~\cite{Oscar, VinVL}. Language modeling can be classified into four categories: (1) LSTM-based approaches with single-layer or two-layer options~\cite{showandtell, bottomup}; (2) CNN-based methods attempting to surpass the fully recurrent paradigm~\cite{CNNdecoder}; (3) Transformer-based fully attentive approaches~\cite{transformerDecoder}; and (4) BERT-like strategies that directly connect visual and textual inputs~\cite{Oscar}.

In recent years, many IC approaches have been proposed. Li et al.~\cite{Oscar} proposed Oscar, a BERT-like architecture that included object tags to ease semantic alignment between images and text. Zhang et al.~\cite{VinVL} built on top of Oscar with VinVL, introducing a new object detector and a modified version of the vision-and-language pre-training objectives. Wang et al.~\cite{OFA} proposed OFA that introduced a unified instruction-based task representation and achieved SOTA performance in IC.

\subsection{Metamorphic Testing}
\label{subsec:MT}
Traditionally, software testing relies on a \textit{test oracle} to check the execution result. The \textit{oracle problem}~\cite{oraclesurvey} refers to the challenge of distinguishing the corresponding desired, correct behavior from potentially incorrect behavior.

Metamorphic testing was introduced by Chen et al. \cite{MTandOracle} as a solution to address the oracle problem in traditional software testing. In metamorphic testing, a \textbf{Metamorphic Relation} (MR) represents the relationship between changes in software inputs and corresponding expected changes in outputs during multiple program executions.

A \textbf{Metamorphic Group} (MG) consists of a group of related test cases, where one is the \textit{source test case} and the others are \textit{follow-up test cases} generated by applying the predefined MR to the source test case. The outputs of MGs are compared to check whether they are consistent with the expected relationships specified in the MR. This comparison of outputs aids in detecting potential faults in the SUT.

\subsection{Testing Approaches for IC Systems}
\label{subsec:ictest}

\begin{figure}[ht]
    \centering
    \includegraphics[width=0.48\textwidth]{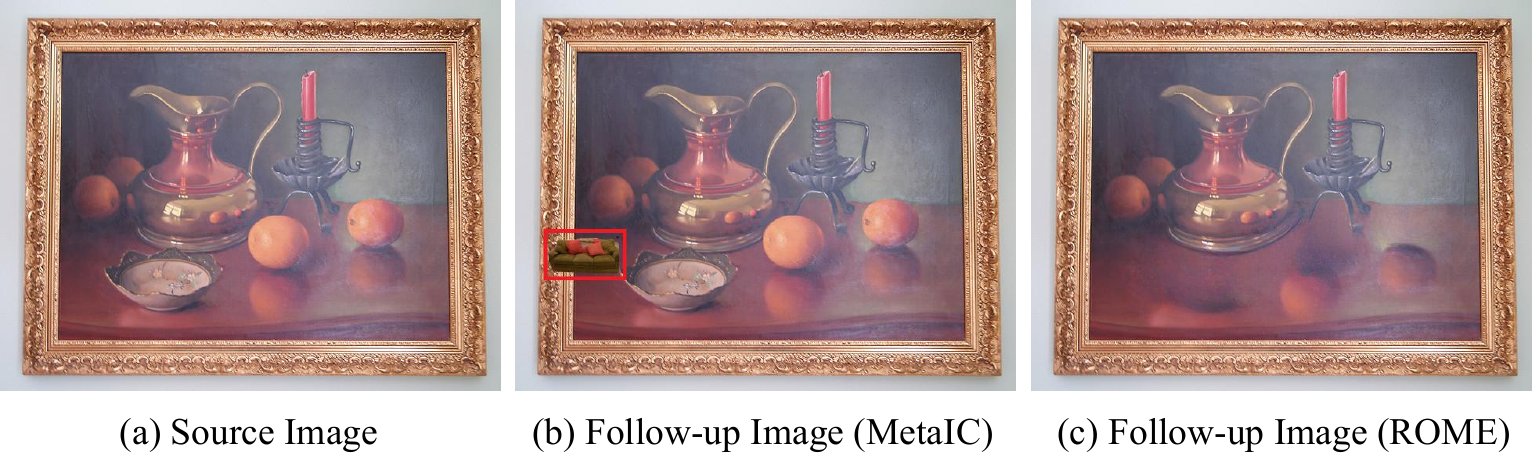}
    \caption{Example of existing testing methods for the IC systems}
    \label{fig:existmethods}
\end{figure}

Currently, there exist two approaches for testing the IC systems. MetaIC~\cite{MetaIC_Pinjia} is the first testing method for IC systems. It is based on the idea that the caption of a background image should undergo a directional change after the insertion of a new object. Specifically, it first constructs an object corpus by extracting objects from existing images using an object segmentation method. Then, as shown in Fig.~\ref{fig:existmethods}(b), it inserts an object from the corpus into the background image using algorithms designed to resize the inserted object and tune its location so as to avoid influencing pre-existing objects. Finally, MetaIC reports a violation if the model outputs for an image pair violate the output relation of MRs. MetaIC outlines two output relations to check violations: (1) The caption generated for the follow-up image should only contain the original objects in the source caption and the inserted object. (2) In the follow-up captions, both the original objects and the inserted objects should be described in appropriate singular-plural form.

It is challenging to keep the realism of the image when inserting objects into it \cite{reality}. Thus, the follow-up images generated by MetaIC may be unnatural~\cite{ROME_Pinjia}. To tackle this issue, Yu et al. have proposed ROME~\cite{ROME_Pinjia}. As shown in Fig.~\ref{fig:existmethods}(c), ROME removes objects from the source images and then uses an image painting technique, i.e., LaMa~\cite{lama}, to generate follow-up images. Similarly, ROME introduces two output relations: (1) In the generated captions, the object set of the follow-up image (image after removal) should be the subset of the object set of the source image. (2) If all the objects of a class are removed, the class should not be included in the generated caption of the follow-up image.

However, both existing methods still suffer from several limitations. We will discuss their limitations in detail in Section~\ref{subsec:motivation}.

\section{Preliminaries}
\label{sec:Preliminaries}

\subsection{Motivation}
\label{subsec:motivation}

As mentioned in Section~\ref{sec:introduction}, current IC system testing methods still manifest several limitations. In this section, we discuss these limitations that motivate our work. In this work, we propose a new testing method called \methodname{} to address them.

\subsubsection{\textbf{Reliance on manually annotated information}} Current IC testing methods mandatorily rely on some annotated information. 
Specifically, MetaIC relies on the \textit{bounding boxes} provided by the dataset to determine the specific locations of all objects in the original image to avoid affecting existing objects in the background image. Such information takes a huge amount of human effort to annotate~\cite{MSCOCO, PASCAL}. However, it is mandatory to select appropriate object insertion positions to generate test cases and conduct testing in MetaIC. Similarly, ROME employs the \textit{corresponding mask} and corresponding \textit{category label} information to identify the positions and categories of all objects in images. This information also requires much laborious annotation. ROME uses this information to identify removal objects from test images and check violations. 

Unlike the above existing works, \methodname{} allows for testing with any unannotated image, and hence can significantly enlarge the available test cases and save labor power. 

\subsubsection{\textbf{Unrealism of generated test images}}
Current IC testing methods have not completely overcome the issue of generating unrealistic images. Specifically, MetaIC inserts new objects into the source images to formulate new follow-up images. However, it can inappropriately regulate the size and position of the inserted objects, leading to unnatural arrangements or inappropriate scaling of objects in the new images~\cite{ROME_Pinjia}. For example, as shown in Fig.~\ref{fig:existmethods}(b), a tiny \textit{sofa} is inserted into the image of a painting with some fruit on a table, which is obviously unrealistic in terms of the object size and surroundings. Meanwhile, ROME may melt an object \textbf{partially} or may not correctly handle \textbf{shadows and reflections}. For example, as shown in Fig.~\ref{fig:existmethods}(c), ROME does not remove the corresponding shadows when melting the oranges, which results in an unrealistic case. We summarize major types of unrealism and give more examples in Section~\ref{subsec:discussreality}\footnote{All examples are generated with the provided scripts by~\cite{MetaIC_Pinjia} and~\cite{ROME_Pinjia}.}.

In contrast, \methodname{} does not artificially manipulate any object in an image. Instead, it adopts reduction-based transformations on the entire image. Therefore, it will preserve the original naturalness of the image, and will not bring in any potential unrealistic factors.

\subsubsection{\textbf{Low eligibility of source test cases}}
In existing methods, only some specific images can be regarded as eligible source test cases for generating follow-up ones. MetaIC proposes the requirement about the available insertion position, object size, and category of the pre-existing objects for the eligible source images. Specifically, if the background image does not provide adequate space for the object insertion or the inserted object still obscures the objects in the source image after several attempts, the test case will not be utilized as an eligible test case. Similarly, ROME only leverages the images that include at least one removable object to perform testing. Based on our experimental findings, among the 300 images available from the PASCAL dataset \cite{PASCAL}, MetaIC and ROME can only utilize 60 and 150 images as the eligible test cases, respectively.

The issue of low eligibility over the available images does not exist in \methodname{} because the transformations in \methodname{} can be generally applied on any image.

\subsection{Challenges}
\label{subsec:challenges}

\subsubsection{\textbf{Localizing objects in captions generated by the IC systems}} As mentioned in Section~\ref{sec:introduction}, to dispense the reliance on annotations, \methodname{} needs to establish correspondence between objects in the source caption and objects in the test image. Thus, the first challenge lies in properly localizing the objects described in the caption. While OD models can be used to locate objects in the image, simply using OD models for localization can be inadequate as it may miss the objects in the caption for three reasons: (1) OD models may miss some objects, resulting in the inability to find positions for specifically described objects. (2) Due to the limitations of current semantic similarity tools, establishing correspondence between the output tags from OD models and the words in captions might be hard. Even if an object is detected by the OD model, its class tag predicted by the OD model may not match the words in the generated captions.
(3) The output of IC systems might contain errors, such as misidentifying a \textit{cat} as a \textit{dog}, making it challenging to ``locate'' the described objects accurately, since OD rarely experiences the same misclassification simultaneously.

\subsubsection{\textbf{Filtering out ambiguous case}
\label{subsubsec:filteringambiguous}}
\methodname{} adopts reduction-based transformations (such as cropping, stretching, etc.), and decides the relation between source and follow-up captions according to whether each object in the source caption is reduced after the transformation. This design considers that the following ambiguous scenario is invalid and should be avoided in the testing: when an object is cut partially, neither classifying it as retained (i.e., to be retained in the follow-up caption) nor disappeared (i.e., to disappear in the follow-up caption) is appropriate. However, reduction-based transformations cannot avoid this ambiguous situation. Therefore, we need a strategy to filter out these ambiguous cases and hence increase the number of valid follow-up inputs.

\subsubsection{\textbf{Boosting the diversity of revealed violations}} Given an image as the source input, it may reveal multiple violations (along with different follow-up inputs). However, it is possible that these violations are quite similar to each other (e.g., they are revealed due to the same missing object). It is generally acknowledged that a good testing method should disclose more distinct failure behaviors. Thus, we should also enhance the diversity of follow-up images to detect more distinct violations~\cite{diversity1, diversity2}. We will discuss distinct violations in Section~\ref{subsec:caseGeneration}.

\section{Method}
\label{sec:Methodology}

\begin{figure*}[ht]
    \centering
    \includegraphics[width=0.87\textwidth]{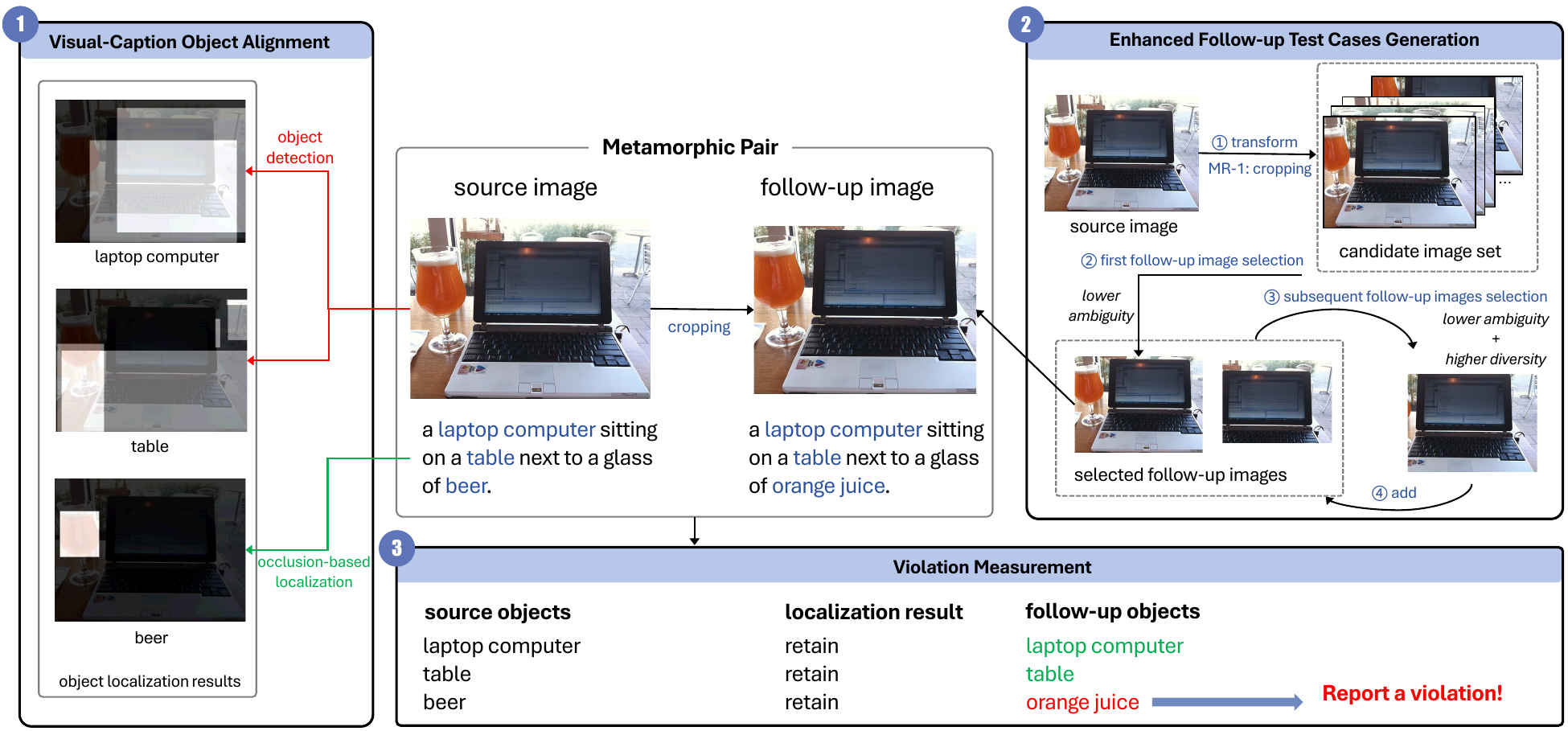}
    \caption{Overview of our method}
    \label{fig:overview}
\end{figure*}

\subsection{Overview of Our Testing Method}
\label{subsec:MethodOverview}

In this section, we introduce our method, \methodname{}. As highlighted in Section~\ref{subsec:challenges}, we face three challenges in this work. Accordingly, we have designed the following modules to address these challenges. As shown in Fig.~\ref{fig:overview}, the overall workflow of \methodname{} can be summed up into the following steps:

\begin{itemize}
    \item \textbf{Visual-Caption Object Alignment.} This module aims to align the objects described in the caption with their positions in the image, in order to address the first challenge. To achieve this, we design a localization method that combines OD and occlusion-based localization.

    \item \textbf{Enhanced Follow-up Test Cases Generation.} We adopt three reduction-based MRs to generate follow-up test images. Each MR is configured with one or several parameters and hence multiple follow-up test cases can be generated from one source test case with different parameter settings. For example, we can generate multiple follow-up images by adopting MR-3 with different rotation angles. However, not all the follow-up test cases have the same contribution to fault detection. So, this module aims to select follow-up test cases with lower ambiguity and higher diversity, in order to tackle the second and the third challenges.

    \item \textbf{Violation Measurement}. After obtaining the source and follow-up outputs, we check them with three detection rules to measure whether there exists a violation by comparing the objects in the source caption and follow-up caption with semantic similarity evaluation.
\end{itemize}

\methodname{} addresses the limitations of the existing methods introduced in Section~\ref{subsec:motivation}. \textit{For the first limitation}, \methodname{} uses a localization method that combines OD and occlusion-based localization to align the objects in the caption with those in the test image. This helps \methodname{} to determine whether objects disappear due to transformation to cut its dependency on pre-annotated labels. The result of RQ1 reported in Section~\ref{subsec:RQ1} will demonstrate the good capability of \methodname{}. \textit{For the second limitation}, we introduce reduction-based transformations that do not artificially manipulate any objects and preserve the realism of generated images. We discuss the scenarios in which existing works lead to the generation of unrealistic images in Section~\ref{subsec:discussreality}, and these scenarios do not occur in the images generated by \methodname{}. \textit{For the last limitation}, we do not set requirements on the eligibility of source test cases. The result of RQ2 reported in Section~\ref{subsec:RQ2} will show that \methodname{} achieves higher eligibility on test images than existing works.

\subsection{Visual-Caption Object Alignment}
\label{subsec:loc}

\begin{figure*} [h]
	\centering
	\subfloat[a \textcolor{red}{laptop computer} sitting on a \textcolor{red}{table} next to a glass of \textcolor{red}{beer}.]{
		\includegraphics[width=1.35 in]{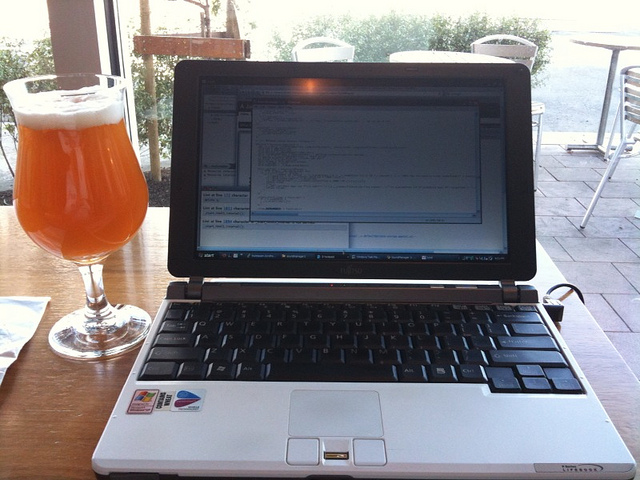}}\hspace*{-\fboxsep}
	\subfloat[Object detection result]{
		\includegraphics[width=1.35 in]{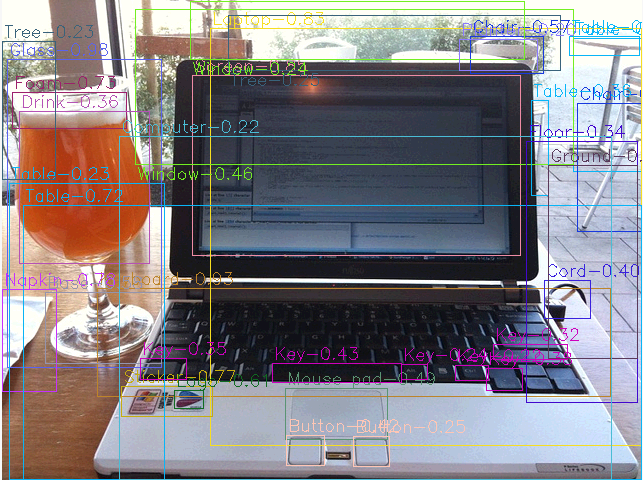}}
    \hspace{0pt}
	\subfloat[Localization result for \textit{laptop computer}]{
		\includegraphics[width=1.35 in]{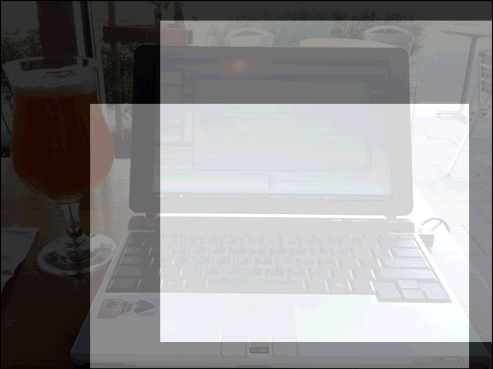}}
	\caption{Example of object detection localization}
	\label{fig:od}
\end{figure*}

This module aims to find the positions of the objects described in the source caption. As discussed in Section~\ref{subsec:challenges}, solely using the OD model is not enough to achieve this goal. 
We employ a combination of the OD model and occlusion-based localization method to achieve the localization. 

\subsubsection{Extracting Objects from Captions} The first step is extracting the objects in the caption. To achieve this, we utilize a Part-Of-Speech (POS) tagging tool 
to extract nouns from the sentence. We conduct two processes on the extracted words. Firstly, we exclude nouns that appear before \textit{``of''}, as these nouns typically function as quantifiers or modifiers. Secondly, noun phrases that are connected are treated as a word-phrase.

\subsubsection{Object Detection} To align an object in the caption with its position in the image, a straightforward method is to obtain all objects in the image and then check whether the object in the caption matches any of these objects detected by OD models. OD models can locate objects in an image and outline their positions and categories using bounding boxes~\cite{ODSurvey}. Therefore, we employ a deep learning-driven OD model~\cite{SGGOD} to correlate the objects in the caption and their positions within the image. As shown in Fig.~\ref{fig:od}, \textit{``laptop computer''} described in the caption matches the output by the OD model. Hence, we can use a bounding box from the output of the OD model as the position of \textit{``laptop computer''} within the image.

As a reminder, the evaluation of whether two objects belong to the same category in this module does not rely on the literal word comparison. Instead, we consider the \textit{semantic similarity}. Specifically, to verify whether two objects belong to the same category, we utilize the Gensim library~\cite{gensim} to calculate the semantic similarity between two words based on the text representations provided by FastText~\cite{fasttext}. FastText is widely employed for various tasks, including text classification, language identification, and calculating the similarity between words. Additionally, 
certain vocabulary terms may share \textit{hyponymy relations}~\cite{wordRelation} (e.g., \textit{man} and \textit{person}), indicating that they represent the same category of objects. FastText might not provide a high similarity score for them. To address this, we also apply WordNet to check if there are mutual hyponymy relationships between two words. WordNet~\cite{wordnet} is a lexical database that provides a hierarchical structure of words, grouping them into sets of synonyms called synsets. This strategy is also employed in all latter components where semantic comparisons are needed.

\subsubsection{Occlusion-based Localization} As mentioned in Section~\ref{subsec:challenges}, solely using the OD model may be not enough to locate the objects. For example, let us consider \textit{``beer''} in the caption of Fig.~\ref{fig:od}(a). In Fig.~\ref{fig:od}(b), we can observe that in cases where the output of the OD model contains a label \textit{``drink''}, it cannot be accurately matched due to the limitation of semantic similarity tools. Moreover, as mentioned in Section~\ref{subsec:challenges}, OD models and IC systems may also produce incorrect outputs, which can also prevent the usage of OD to ``locate'' the described objects. To tackle this issue, we further propose an occlusion-based localization method.

Some DL model interpretation methods find key features by iteratively occluding areas in the image~\cite{RISE, DRISE}. Inspired by them, we locate the described objects with occlusion. However, these methods use pixel-level occlusion for interpretation, which does not effectively meet our purpose of aligning objects in the caption and the test image. Thus, we design the \textit{object-level occlusion} to realize our goal. Specifically, we utilize the bounding boxes obtained from the preceding object detection step. This allows us to occlude complete objects at a time, rather than dealing with sets of pixels.

\begin{figure} [ht]
	\centering
	\subfloat[Blurring]{
		\includegraphics[width=1.25 in]{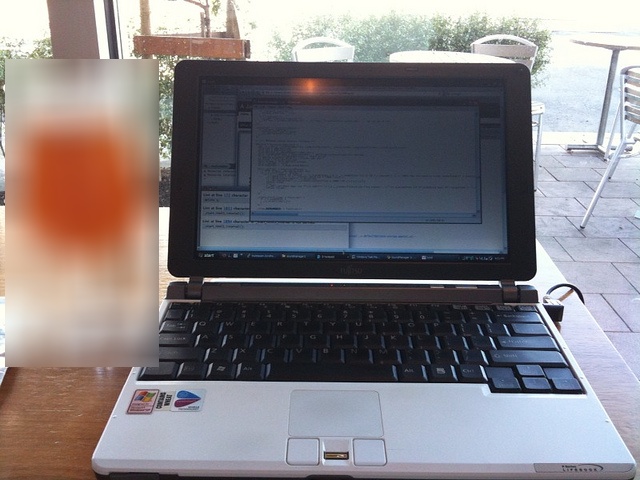}}
	\subfloat[Black block filling]{
		\includegraphics[width=1.25 in]{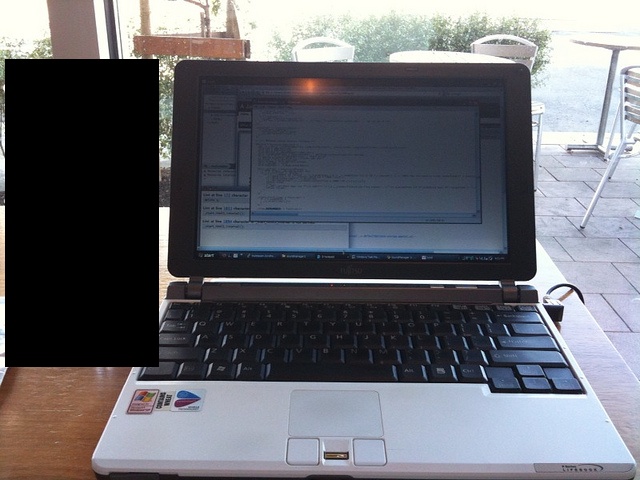}}
	\quad
	\subfloat[Image inpainting]{
		\includegraphics[width=1.25 in]{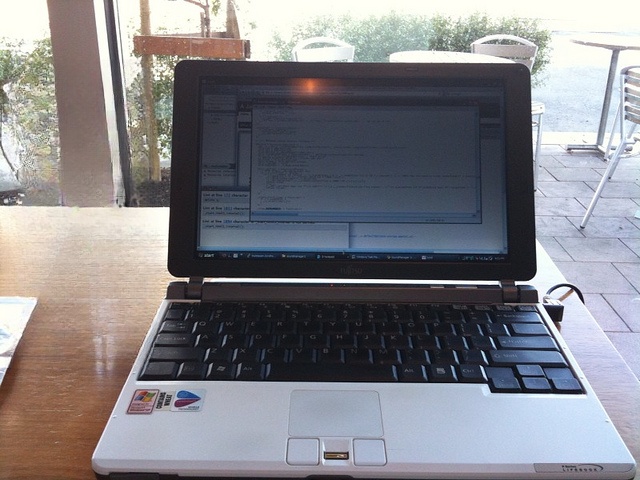}}
	\subfloat[Localization result for \textit{beer}]{
		\includegraphics[width=1.25 in]{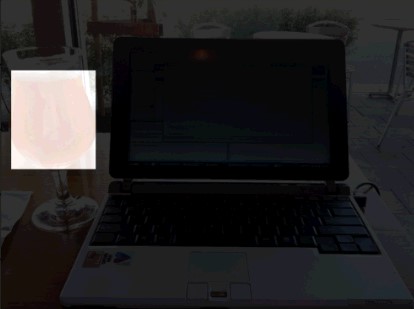}}
	\caption{Example of occlusion-based localization}
	\label{occ}
\end{figure}

The occlusion-based localization method locates the position of an object by first occluding a region in a test image and next checking if the corresponding caption of the occluded image no longer includes the object. To reduce randomness, we incorporate three different occlusion methods. We only consider a region as the position of an object when the object is absent from the captions for \textbf{all three occluded versions}. As illustrated in Fig.~\ref{occ}, three occlusion methods are as follows:

\begin{itemize}
    \item \textbf{Blurring:} The region to be occluded applies the Gaussian blur algorithm.

    \item \textbf{Black block filling:} The region to be occluded is filled with all pixels set to 0 (black).

    \item \textbf{Image inpainting:} For the region requiring occlusion, we remove its content and perform image inpainting to fill the region. We utilize the popular image inpainting model, Lama~\cite{lama}.\footnote{As a reminder, the realism of occluded images is not a problem in our method, since we do not use it to generate any follow-up test images.}
\end{itemize}

\begin{algorithm}
    \caption{Occlusion-based Localization Algorithm}\label{alg:occ}
    \SetKwInOut{Data}{Input}
    \SetKwInOut{Result}{Output}
    \Data{$o$: object in the caption to be located;\\
    $I_s$: source image;\\
    $BBoxes$: object in  bounding boxes of $I_s$;\\
    $IC$: SUT}
    \Result{$M_{loc}$: localization result of $o$}
    $regions \xleftarrow{}$ emptyList\;
    \For{$box$ in $BBoxes$} {
        $caption_1$ = $IC$(blur($I_s, box$))\;
        $caption_2$ = $IC$(blackFilling($I_s, box$))\;
        $caption_3$ = $IC$(inpainting($I_s, box$))\;
        \If{$o$ not in $caption_1$ \textbf{AND} $o$ not in $caption_2$ \textbf{AND} $o$ not in $caption_3$}{
            $regions$.add($box$)\;
        }
    }
    $regions$ = removeInclusion($regions$)\;
    \For{$region$ in $regions$}{
        $M_{loc}$.add($region$)\;
    }
    \Return{$M_{loc}$}\;
\end{algorithm}

The process of our occlusion localization method is described in Algorithm~\ref{alg:occ}. We denote $o$ as the object to be located, which is described in the caption of source image $I_s$. Firstly, in lines 2-5, we use the bounding boxes predicted by the OD model and occlude one region at a time in the source image to generate a series of occluded images. After that, we feed these occluded images into the SUT to obtain caption\textsubscript{1}, caption\textsubscript{2}, and caption\textsubscript{3}, respectively. Secondly, in lines 6-7, we check whether these captions contain $o$. Only if $o$ is absent in all caption\textsubscript{1}, caption\textsubscript{2}, and caption\textsubscript{3}, we add the bounding box of the occluded region to the $regions$ of the current object. Otherwise, if one or more captions of occluded images contain $o$, that region will not be considered as the position of $o$. Thirdly, in line 10, if two occlusion regions have an \textbf{inclusion} relationship, the smaller one is considered to be more likely to represent the precise location of the object. Therefore, we remove the larger box if there is such a relationship. Finally, we use boxes in $regions$ to construct a matrix that indicates the position of this object in the image. By utilizing occlusion-based localization, we can locate the objects in the source caption and align them with objects in the image, thus effectively supplementing the missing objects during OD localization.

\subsection{Metamorphic Relations}
\label{subsec:MRs}

Typically, an MR consists of two essential components: \textbf{input transformation} and \textbf{output relation}~\cite{MTsurvey}. In the context of IC testing, input transformation is responsible for generating follow-up images from source images using transformation methods; while output relation defines the expected relationship that should hold between the captions that the IC system generates for the source and follow-up images.

\begin{figure} [ht]
	\centering
	\subfloat[source image]{
		\includegraphics[width=1.5 in]{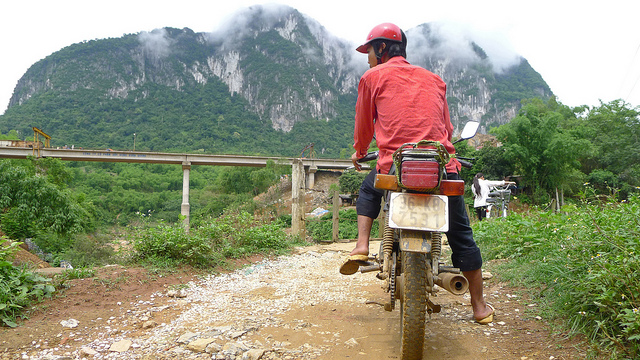}}
	\subfloat[cropping (MR-1)]{
		\includegraphics[width=1.5 in]{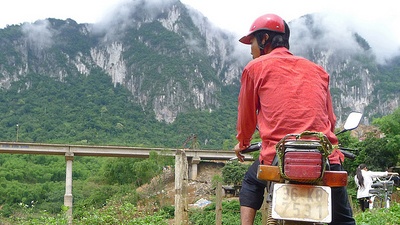}}
	\quad
	\subfloat[stretching (MR-2)]{
		\includegraphics[width=1.5 in]{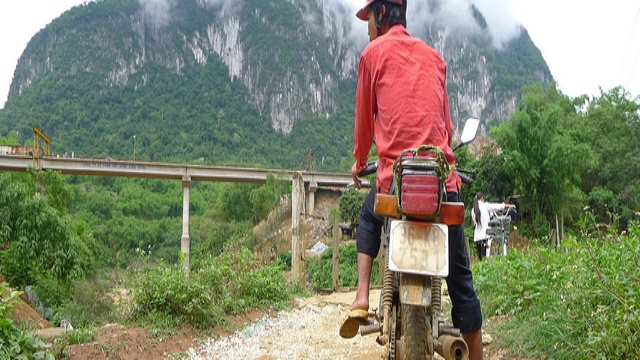}}
	\subfloat[rotation (MR-3)]{
		\includegraphics[width=1.5 in]{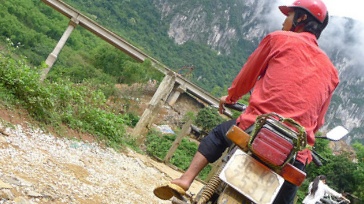}}
	\caption{Example of our input transformation methods}
	\label{inputTransformation}
\end{figure}

\textbf{Input Transformation in \methodname{}}. As mentioned in Section~\ref{sec:introduction}, \methodname{} avoids manipulating particular objects. Instead, it performs transformations on the entire image. Our transformations follow a reduction-based paradigm, which may reduce information in the image. In this paper, we use three reduction-based transformations as representative, i.e., cropping, stretching, and rotation. Fig.~\ref{inputTransformation} illustrates our transformations.

\begin{itemize}
    \item \textbf{Cropping (MR-1):} Cut out a portion of the image without altering its proportions. The cropping process is determined by two parameters: the top-left point and the bottom-right point.
    
    \item \textbf{Stretching (MR-2):} Elongate the image along a fixed axis and discard the portion that exceeds the image boundaries. The stretching process is also determined by two parameters: the two points in the axis.

    \item \textbf{Rotation (MR-3):} Rotate the image by a specific angle. It is determined by a single parameter: the angle of rotation. Note that rotation often introduces dark borders encircling the image, which makes the image devoid of realism. Thus, we perform an additional cropping step to eliminate these dark borders.
\end{itemize}

To avoid reducing too much information in the test images, we devise guidelines for these three input transformation methods. For the cropping transformation in MR-1, the cropped image must be larger than the average size of the objects detected by the OD model. This ensures that the image has enough space to accommodate at least one object. For the stretching transformation in MR-2, the retained proportion of the image must be at least 0.6. For the rotation transformation in MR-3, the rotation angle is limited to [-30, 30] degrees\footnote{These settings are based on our observations from our preliminary exploratory experiments, which also basically comply with common sense.}. 

\textbf{Output Relation In \methodname{}}. The expected output relation for all three reduction-based transformations is the same. In the \textit{Visual-Caption Alignment} module, we obtain the localization results of described objects in the source caption. These results serve as the basis for determining the output relation, that is, whether the described objects in the source caption should retain or disappear after the transformation.

As depicted in Fig.~\ref{occ}(d), the localization results can be represented as a matrix $M_{loc}$ of the same dimensions as the image. We can also construct an equally sized $M_{tran}$ after the source image has been transformed. In $M_{tran}$, \textit{1} indicates that the corresponding pixel in the follow-up image still \textbf{retains}, while \textit{0} indicates the absence of this pixel. The \textbf{retain ratio} $Ratio_r$ of an object after transformation is calculated as:

\begin{equation}
Ratio_r=\frac{sum(M_{loc}) \times sum(M_{tran})}{sum(M_{loc})} 
\end{equation}

According to the $Ratio_r$ after transformation, objects in the caption can be classified into the following three types.

\begin{align*}
    \textbf{Retain}&: Ratio_r \in \left[T_{up},1\right] \\
    \textbf{Ambiguous}&: Ratio_r \in \left(T_{down},T_{up}\right) \\
    \textbf{Disappear}&: Ratio_r \in \left[0,T_{down}\right]
\end{align*}

$T_{down}$ and $T_{up}$ are predefined thresholds. We have made some preliminary observations and have explored various threshold combinations, then we can finalize our threshold values: 0.2 for $T_{down}$ and 0.9 for $T_{up}$.

The three types of objects in the MR transformation are treated differently. For the \textit{retained} objects, we expect them to continue appearing in the follow-up caption. For the \textit{disappeared} objects, we expect them to be absent in the follow-up caption. As mentioned in Section~\ref{subsubsec:filteringambiguous}, we do not check the output relation on the ambiguous objects. Note that this is a brief overview of the output relation in \methodname{}, please refer to Section~\ref{subsec:violationCheck} for the detailed output checking procedure.

Since we cannot define the violation behavior on the ambiguous objects and thus such objects are helpless to detect the SUT failures, we would like to \textbf{minimize the occurrence of ambiguous objects as much as possible} to enhance the effectiveness of the test suites. We design the \textit{Enhanced Follow-up Test Cases Generation} module (introduced in Section~\ref{subsec:caseGeneration}) to reduce the occurrence of ambiguous cases. 

\subsection{Enhanced Follow-up Test Cases Generation}
\label{subsec:caseGeneration}

\begin{figure*}[h!]
    \centering
    \includegraphics[width=0.81\textwidth]{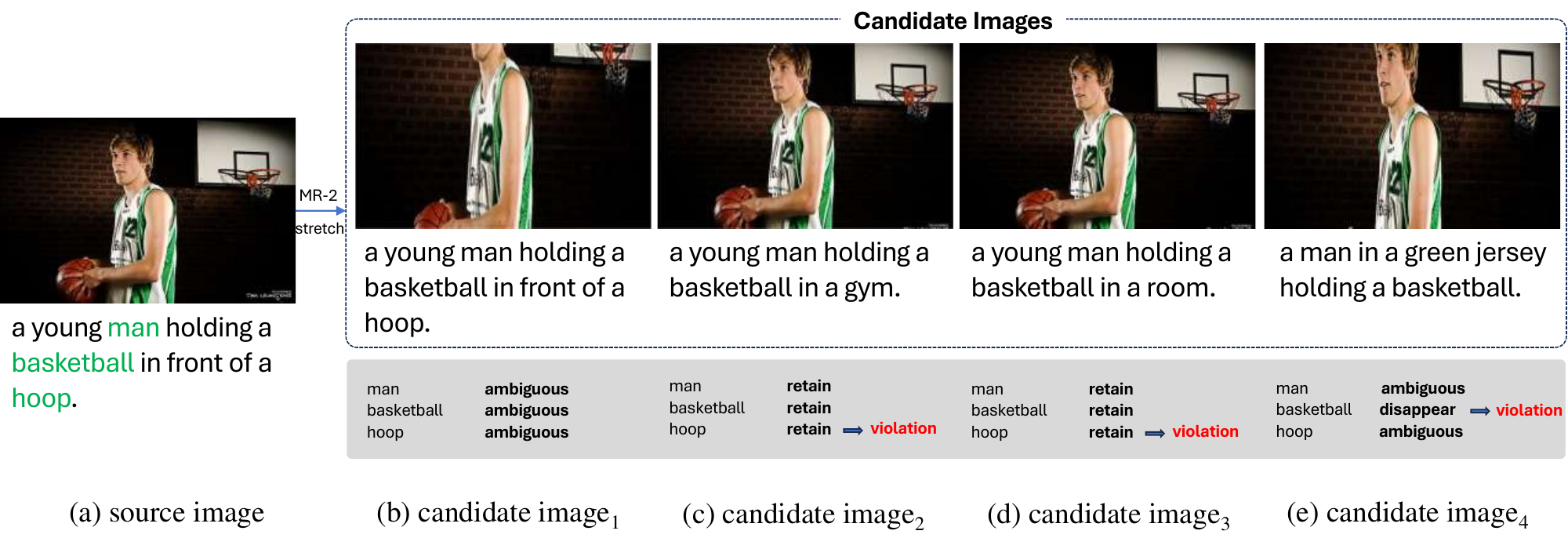}
    \caption{Enhanced follow-up test cases generation}
    \label{fig:enhance_followup}
\end{figure*}

This module aims to tackle the second and third challenges in Section~\ref{subsec:challenges}.

If an object disappears partially after transformations, we are hardly able to detect violations related to such uncertain objects. In particular, those candidate images whose objects described in the source caption are all ambiguous will not be considered as \textbf{valid cases}. For example, in Fig.~\ref{fig:enhance_followup}(c) and (d), all objects (i.e., ``\textit{man}'', ``\textit{basketball}'', and ``\textit{hoop}'') in the source caption are recognized as \textbf{retained} in the corresponding follow-up images. Obviously, these two cases are suitable for our testing, since it is reasonable to expect all these three objects to appear in the follow-up captions. In this example, captions in these two follow-up cases fail to include ``\textit{hoop}'' (as well as tokens with similar semantics), and hence reveal violations. In contrast, all three objects in the source caption have been recognized as \textbf{ambiguous} in Fig.~\ref{fig:enhance_followup}(b), which brings uncertainty in deciding the output relation, and hence this follow-up image \textbf{should not be included} during testing.

Meanwhile, when generating multiple follow-up images, the diversity of the revealed violations should be considered. Specifically, it is \textbf{much better} to have a number of violations revealed on \textbf{distinct objects} than on the same object. For example, the violations in Fig.~\ref{fig:enhance_followup}(c) and Fig.~\ref{fig:enhance_followup}(d) cover the same object \textit{``hoop''}. Therefore, in the selection of subsequent follow-up images, we should preferably consider violations that \textbf{cover more distinct objects}. In this example, the violation detected in Fig.~\ref{fig:enhance_followup}(e) covers a distinct object \textit{``basketball''}. Obviously, having (c) and (e) (or (d) and (e)) as our follow-up test cases is better than having (c) and (d) since the former option reveals more distinct violations than the latter one.

Thus, this module aims to generate test images with \textbf{lower ambiguity} \textbf{and higher diversity}. However, simply applying the transformation methods with fixed parameters cannot realize this goal. To tackle this issue, we propose a method to select images from multiple candidates to achieve two primary goals: 1) reduce the ambiguous cases, and 2) promote the diversity of generated cases.

\subsubsection{Candidate Image Set Generation}
At first, we generate a number of candidate images for each source image by altering the parameters of transformations. The candidate image set serves as the pool for generating follow-up images. We design different strategies to generate the candidate image set for one-parameter and multi-parameter transformation methods.

For the one-parameter transformations, such as rotation, we directly enumerate the parameter within its variable range \textbf{uniformly}. For instance, in the case of rotation, we uniformly sample angles within the interval of [-30, 30] degrees at intervals of 1 degree.

For the multiple-parameter transformations, such as cropping and stretching, we design another strategy. To create the candidate image set for these transformations, we uniformly select and establish one of the parameters and iterate through the feasible range of the other parameters. For instance, we uniformly select ten top-left points and then iterate feasible bottom-right points to generate cropping images.

\subsubsection{The First Follow-up Image Selection}

When selecting the first follow-up image from the candidate image set, we do not need to regard its diversity. Thus, the primary concern is \textbf{to mitigate the occurrence of ambiguous objects}.

For each image in the candidate image set, we check if the objects described in the source output retain, become ambiguous, or disappear based on the localization results.

We denote $N_{retain}$, $N_{ambiguous}$, and $N_{disappear}$ as the numbers of retained, ambiguous, and disappeared objects in the source caption, respectively. To minimize the number of ambiguous objects, the value of $N_{ambiguous}$ should be as small as possible. 
Meanwhile, we find that the images with very little difference from the source test case usually have a small ambiguous score. Consider that dissimilar test cases tend to have better fault-revealing ability~\cite{adaptiveMT, ause23qaasker+,ase22MPPrioritize}. Therefore, we also introduce $N_{disappear}$ in prioritization to select the images with both less ambiguous objects and sufficient perturbation as the follow-up test cases. The final score is calculated as:

\begin{equation}
    Score_{ambiguity}^{i}=N_{disappear}^{i} - N_{ambiguous}^{i}
\end{equation}

After computing $Score_{ambiguity}$ for all candidate images, we select the one with the highest score as the first follow-up image corresponding to the current source image.

\subsubsection{Subsequent Follow-up Images Selection}

When selecting the subsequent follow-up images, it is important to \textbf{maintain diversity among the follow-up images} in addition to reducing the occurrence of ambiguous cases. 

For this goal, we first define the method to calculate the difference between two images. Specifically, for two candidate images $i$ and $i'$, we obtain $M_{tran}^{i}$ and $M_{tran}^{i'}$ based on the method described in Section~\ref{subsec:MRs}. Then, we calculate the difference between $i$ and $i'$ is as follows:

\begin{equation}
    Diff_{i,i'} = 1 - \frac{M_{tran}^{i} \cap M_{tran}^{i'}}{M_{tran}^{i} \cup M_{tran}^{i'}}
\end{equation}

We denote the already selected images as set $\mathbb{I}_{selected}$. Then, for the images in the candidate set, we calculate their diversity score as follows:

\begin{equation}
    Score_{diversity}^{i}=min(Diff_{i,c}\ |\  c \in \mathbb{I}_{selected})
\end{equation}

To balance $Score_{ambiguity}$ and $Score_{diversity}$, we combine the two scores to calculate the final $Score$ for each candidate image. In order to scale the two scores into the same data magnitude to ensure consistent contributions to the final $Score$, we first normalize them \textbf{separately}. The normalization process is identical for both scores. To illustrate, we take the calculation for $Score_{ambiguity}$ for example. Firstly, we compute the $Score_{ambiguity}$ for \textbf{all candidate images} and calculate the maximum score ($max\_score$) and minimum score ($min\_score$) among them. Then, $Score_{ambiguity}^{i}$ can be scaled to the range of [0,1] as follows:

\begin{equation}
\label{eq:norm}
    norm(Score_{ambiguity}^{i}) = \frac{Score_{ambiguity}^{i} - min\_score}{max\_score - min\_score}
\end{equation}

Finally, we get the final $Score$ for each candidate image as:

\begin{equation}
    Score^i=norm(Score_{ambiguity}^{i}) + norm(Score_{diversity}^{i})
\end{equation}

We continue the process by selecting the subsequent follow-up images {one by one}. During each iteration, we select the image with the highest $Score$ from the candidate image set and then add the image to the selected set. This process continues until the desired number of selected images is reached or no more candidate images are available.

\subsection{Violation Measurement}
\label{subsec:violationCheck}

By generating multiple follow-up images for each source image, we pair them with their corresponding source images to form image pairs ($I_s, I_f$). Let us denote the corresponding captions as ($C_s, C_f$). For each of these pairs, we utilize the object localization results obtained in Section~\ref{subsec:loc} and design the following rules of expected behaviors. If a pair violates any of these rules, a violation will be reported, which indicates that errors have occurred in at least one of the source outputs or the follow-up output. For simplicity in the following introduction, we use $\mathbb{S}_{s}$ and $\mathbb{S}_{f}$ to refer to the object sets for $C_s$ and $C_f$, respectively. For an MG, we can classify $\mathbb{S}_{s}$ into three sets based on the localization results of each object and the transformation applied in the follow-up: $\mathbb{S}_s^{retain}$ for objects that retain after transformation, $\mathbb{S}_s^{ambiguity}$ for objects that are ambiguity after transformation, and $\mathbb{S}_s^{disappear}$ for objects that disappear after transformation.

\textbf{Rule 1:} If an object $o$ described in the source caption continues retaining after transformation, it should be described in the follow-up caption. Rule 1 is formally defined as follows:

\begin{equation}
    o \in \mathbb{S}_{f}\ for\ \forall o \in \mathbb{S}_s^{retain}
\end{equation}

\textbf{Rule 2:} If an object $o$ described in the source caption disappears after transformation, it should not be described in the follow-up caption. Rule 2 is formally defined as follows:

\begin{equation}
    o \notin \mathbb{S}_{f}\ for\ \forall o \in \mathbb{S}_s^{disappear}
\end{equation}

\textbf{Rule 3:} If \textbf{all objects} described in the source caption are retained in the image after transformation, the content of the source image and the follow-up image is \textbf{equivalent} as no salient object has been removed. We subsequently verify if $\mathbb{S}_{s}$ is equivalent to $\mathbb{S}_{f}$. Rule 3 is formally defined as follows:

\begin{equation}
    \mathbb{S}_{s}=\mathbb{S}_{f}, \ if \ \mathbb{S}_s^{ambiguity}=\emptyset \ and\  \mathbb{S}_s^{disappear}=\emptyset
\end{equation}

For a given MG, we start by obtaining the localization results of all objects in the source caption. This helps \methodname{} determine whether each object disappears after transformation. Then, we evaluate whether an MG violates the predefined rules above to reveal if any violation occurs.

As a reminder, in the above measurement, we compare two words by following the same semantic-aware method as the \textit{Visual-Caption Object Alignment} module in Section~\ref{subsec:loc}.

\section{Experiment Setup}
\label{sec:exp}

\subsection{Research Questions}
We evaluated our approach with the following research questions: 

\textbf{\textit{RQ1: How effective is our approach at detecting violations in IC systems?}} This RQ aims to investigate if the violations reported by \methodname{} indeed represent actual violations using images {without annotations}. We also analyzed the types of errors that our method can detect.

\textbf{\textit{RQ2: How does the eligibility of test cases of our approach compare to existing methods?}} This RQ compares the eligibility on test images of our approach and existing methods which depend on the annotated information.

\textbf{\textit{RQ3: What is the quality of follow-up test cases generated by the Enhanced Follow-up Test Cases Generation module?}} 
As introduced in Section~\ref{subsec:motivation}, our method is good at generating effective test cases. This RQ studies the quality of generated test images in terms of their ambiguity and diversity. We compared the quality of the test cases generated by \methodname{} and the methods without considering the ambiguity and diversity of test cases. 

\textbf{\textit{RQ4: What is the impact of various components in \methodname{} on violation detection using generated MGs?}} In this RQ, we set up an ablation study to investigate the impact of various components of \methodname{} on its final fault detection capabilities. Specifically, we ablated the \textit{object detection localization} component, the \textit{occlusion-based localization} component, and the \textit{semantic similarity} component, respectively.

\textbf{\textit{RQ5: How effective is our approach on different models, datasets, and transformation methods?}} This RQ aims to explore the performance of the SUTs in various scenarios by comparing the behaviors during the testing phase.

\subsection{Evaluation Metrics}
We used four common metrics, \textbf{Accuracy, Precision, Recall}, and \textbf{F1-score}, to measure the quality of the reported violations~\cite{CAT}. These metrics are calculated by determining the number of True Positives (TP), False Negatives (FN), True Negatives (TN), and False Positives (FP). In the context of MT, we consider a reported violation as a ``\textit{positive}''. 
If the source and follow-up outputs in an MG violate the expected output relation in the MR and the testing method reports a violation correspondingly, a TP is recorded. If the method fails to report the violation, an FN is recorded. Similarly, if the two outputs follow the MR and the testing method reports no violation, a TN is recorded. Otherwise, an FP is recorded.

To calculate the above metrics and then \textbf{explore the performance of our approach}, we recruited three graduate students who were proficient in English and conducted a manual inspection process to identify \textit{errors} and \textit{the types of TPs}.

Firstly, the inspectors independently determined if the objects in each caption generated by the IC systems matched the content of the corresponding images, and \textbf{recorded an error} if the objects did not. Then, inconsistencies in the recorded errors among the three inspection results were identified again. Specifically, the inspectors need to discuss the inconsistent inspection results and finally reach an agreement.

After the inspectors identified the errors in MGs, we can get the TPs reported by \methodname{}. Then, the inspectors \textbf{categorized the TPs} into three major types (the detailed definition of error types can be found in the second part of Section~\ref{subsec:Test objects}). Similarly, they also need to review and discuss the classification results and reach an agreement.

As a reminder, during the whole testing process, \methodname{} \textbf{automatically} compares the source and follow-up output captions against the MRs to detect and report violations. We perform a manual inspection only for the evaluation of our approach in the experimental analysis.

\subsection{Test Cases}

Although our approach can be applied on any unannotated images, existing methods~\cite{MetaIC_Pinjia, ROME_Pinjia} require images with annotated information. Thus, we used images from two annotated datasets as the source test cases in our evaluation. MSCOCO~\cite{MSCOCO} is the most popular dataset in the field of IC. Meanwhile, PASCAL~\cite{PASCAL} is also a widely-used image dataset that supplies object labels for each image. For each dataset, we randomly selected 300 images as the source cases due to the time cost. Then, we used our three MRs to generate the corresponding follow-up test cases for every source test case. In the experiments of this paper, we set the number of follow-up images generated for each source image to 3.

\subsection{Test Objects}
\label{subsec:Test objects}

We selected five popular IC systems as our test objects \cite{ROME_Pinjia}. Four of them are typical open-source IC models, i.e., Show and Tell~\cite{showandtell}, Oscar~\cite{Oscar}, VinVL~\cite{VinVL}, and OFA~\cite{OFA}. The other one is the commercial Microsoft Azure IC API~\cite{azure}. For Show and Tell, VinVL, and OFA, we used the well-trained models provided by their authors. For Oscar, we fine-tuned the pre-trained model using the instructions from the repository and reproduced the metrics mentioned in the paper. 

\section{Results and Analysis}
\label{sec:Results}

\subsection{Overall Effectiveness (RQ1)}
\label{subsec:RQ1}

\begin{table*}[h!]
\scriptsize
\centering
\caption{Validity of Violations Reported by \methodname{} on Different Dataset and IC Systems}
\begin{tabular}{ccccccccccccc}
\Xhline{0.8px}
\textbf{Dataset} & \textbf{Level} & \textbf{SUT} & \textbf{MR} & \textbf{TP} & \textbf{FN} & \textbf{FP} & \textbf{TN} & \textbf{Accuracy} & \textbf{Precision} & \textbf{Recall} & \textbf{F1-score} \\
\hline
\multirow{30}{*}{MSCOCO} & \multirow{15}{*}{Object} & \multirow{3}{*}{Show and Tell} & MR-1 & 473 & 160 & 56 & 2789 & 93.8\% & 89.4\% & 74.7\% & 81.4\% \\
                         &                          &                                & MR-2 & 597 & 160 & 70 & 3214 & 94.3\% & 89.5\% & 78.9\% & 83.8\% \\
                         &                          &                                & MR-3 & 561 & 163 & 46 & 3447 & 95.0\% & 92.4\% & 77.5\% & 84.3\% \\
\cline{3-12}
                         &                          & \multirow{3}{*}{Oscar}         & MR-1 & 398 & 142 & 57 & 3127 & 94.7\% & 87.5\% & 73.7\% & 80.0\% \\
                         &                          &                                & MR-2 & 577 & 149 & 77 & 3555 & 94.8\% & 88.2\% & 79.5\% & 83.6\% \\
                         &                          &                                & MR-3 & 593 & 119 & 68 & 3807 & 95.9\% & 89.7\% & 83.3\% & 86.4\% \\
\cline{3-12}
                         &                          & \multirow{3}{*}{VinVL}         & MR-1 & 304 & 57 & 67 & 3286 & 96.7\% & 81.9\% & 84.2\% & 83.1\% \\
                         &                          &                                & MR-2 & 474 & 51 & 67 & 3847 & 97.3\% & 87.6\% & 90.3\% & 88.9\% \\
                         &                          &                                & MR-3 & 452 & 77 & 50 & 4101 & 97.3\% & 90.0\% & 85.4\% & 87.7\% \\
\cline{3-12}
                         &                          & \multirow{3}{*}{OFA}           & MR-1 & 218 & 87 & 70 & 3379 & 95.8\% & 75.7\% & 71.5\% & 73.5\% \\
                         &                          &                                & MR-2 & 306 & 59 & 80 & 4077 & 96.9\% & 79.3\% & 83.8\% & 81.5\% \\
                         &                          &                                & MR-3 & 295 & 67 & 54 & 4344 & 97.5\% & 84.5\% & 81.5\% & 83.0\% \\
\cline{3-12}
                         &                          & \multirow{3}{*}{Azure}         & MR-1 & 381 & 65 & 76 & 2560 & 95.4\% & 83.4\% & 85.4\% & 84.4\% \\
                         &                          &                                & MR-2 & 543 & 78 & 61 & 3003 & 96.2\% & 89.9\% & 87.4\% & 88.7\% \\
                         &                          &                                & MR-3 & 578 & 75 & 61 & 3066 & 96.4\% & 90.5\% & 88.5\% & 89.5\% \\
\cline{2-12}
                         & \multirow{15}{*}{Case}   & \multirow{3}{*}{Show and Tell} & MR-1 & 292 & 86 & 23 & 499 & 87.9\% & 92.7\% & 77.2\% & 84.3\% \\
                         &                          &                                & MR-2 & 320 & 54 & 26 & 500 & 91.1\% & 92.5\% & 85.6\% & 88.9\% \\
                         &                          &                                & MR-3 & 302 & 61 & 8 & 529 & 92.3\% & 97.4\% & 83.2\% & 89.7\% \\
\cline{3-12}
                         &                          & \multirow{3}{*}{Oscar}         & MR-1 & 248 & 85 & 27 & 540 & 87.6\% & 90.2\% & 74.5\% & 81.6\% \\
                         &                          &                                & MR-2 & 332 & 52 & 31 & 485 & 90.8\% & 91.5\% & 86.5\% & 88.9\% \\
                         &                          &                                & MR-3 & 329 & 37 & 24 & 510 & 93.2\% & 93.2\% & 89.9\% & 91.5\% \\
\cline{3-12}
                         &                          & \multirow{3}{*}{VinVL}         & MR-1 & 200 & 44 & 41 & 615 & 90.6\% & 83.0\% & 82.0\% & 82.5\% \\
                         &                          &                                & MR-2 & 277 & 30 & 28 & 565 & 93.6\% & 90.8\% & 90.2\% & 90.5\% \\
                         &                          &                                & MR-3 & 263 & 29 & 13 & 595 & 95.3\% & 95.3\% & 90.1\% & 92.6\% \\
\cline{3-12}
                         &                          & \multirow{3}{*}{OFA}           & MR-1 & 169 & 59 & 39 & 633 & 89.1\% & 81.3\% & 74.1\% & 77.5\% \\
                         &                          &                                & MR-2 & 211 & 22 & 40 & 627 & 93.1\% & 84.1\% & 90.6\% & 87.2\% \\
                         &                          &                                & MR-3 & 188 & 24 & 22 & 666 & 94.9\% & 89.5\% & 88.7\% & 89.1\% \\
\cline{3-12}
                         &                          & \multirow{3}{*}{Azure}         & MR-1 & 222 & 35 & 31 & 612 & 92.7\% & 87.7\% & 86.4\% & 87.1\% \\
                         &                          &                                & MR-2 & 277 & 23 & 27 & 573 & 94.4\% & 91.1\% & 92.3\% & 91.7\% \\
                         &                          &                                & MR-3 & 296 & 30 & 21 & 553 & 94.3\% & 93.4\% & 90.8\% & 92.1\% \\
\Xhline{0.8px}
\multirow{30}{*}{PASCAL} & \multirow{15}{*}{Object} & \multirow{3}{*}{Show and Tell} & MR-1 & 566 & 231 & 71 & 2627 & 91.4\% & 88.9\% & 71.0\% & 78.9\%  \\
                         &                          &                                & MR-2 & 683 & 202 & 88 & 2925 & 92.6\% & 88.6\% & 77.2\% & 82.5\%  \\
                         &                          &                                & MR-3 & 641 & 190 & 70 & 3171 & 93.6\% & 90.2\% & 77.1\% & 83.1\%  \\
\cline{3-12}
                         &                          & \multirow{3}{*}{Oscar}         & MR-1 & 506 & 228 & 54 & 2823 & 92.2\% & 90.4\% & 68.9\% & 78.2\%  \\
                         &                          &                                & MR-2 & 694 & 219 & 76 & 3096 & 92.8\% & 90.1\% & 76.0\% & 82.5\%  \\
                         &                          &                                & MR-3 & 726 & 225 & 63 & 3269 & 93.3\% & 92.0\% & 76.3\% & 83.4\%  \\
\cline{3-12}

                         &                          & \multirow{3}{*}{VinVL}         & MR-1 & 317 & 76  & 38 & 3082 & 96.8\% & 89.3\% & 80.7\% & 84.8\%  \\
                         &                          &                                & MR-2 & 458 & 63  & 63 & 3486 & 96.9\% & 87.9\% & 87.9\% & 87.9\%  \\
                         &                          &                                & MR-3 & 439 & 55  & 52 & 3697 & 97.5\% & 89.4\% & 88.9\% & 89.1\%  \\
\cline{3-12}
                         &                          & \multirow{3}{*}{OFA}           & MR-1 & 288 & 63  & 54 & 3253 & 96.8\% & 84.2\% & 82.1\% & 83.1\%  \\
                         &                          &                                & MR-2 & 321 & 49  & 54 & 3784 & 97.6\% & 85.6\% & 86.8\% & 86.2\%  \\
                         &                          &                                & MR-3 & 342 & 57  & 54 & 3923 & 97.5\% & 86.4\% & 85.7\% & 86.0\%  \\
\cline{3-12}
                         &                          & \multirow{3}{*}{Azure}         & MR-1 & 427 & 49  & 50 & 2639 & 96.9\% & 89.5\% & 89.7\% & 89.6\%  \\
                         &                          &                                & MR-2 & 597 & 62  & 68 & 2831 & 96.3\% & 89.8\% & 90.6\% & 90.2\%  \\
                         &                          &                                & MR-3 & 589 & 61  & 50 & 2999 & 97.0\% & 92.2\% & 90.6\% & 91.4\%  \\
\cline{2-12}
                         & \multirow{15}{*}{Case}   & \multirow{3}{*}{Show and Tell} & MR-1 & 322 & 109 & 34 & 435 & 84.1\% & 90.4\% & 74.7\% & 81.8\% \\
                         &                          &                                & MR-2 & 362 & 72 & 34 & 432 & 88.2\% & 91.4\% & 83.4\% & 87.2\% \\
                         &                          &                                & MR-3 & 320 & 65 & 25 & 490 & 90.0\% & 92.8\% & 83.1\% & 87.7\% \\
\cline{3-12}
                         &                          & \multirow{3}{*}{Oscar}         & MR-1 & 297 & 109 & 27 & 467  & 84.9\% & 91.7\% & 73.2\% & 81.4\%  \\
                         &                          &                                & MR-2 & 381 & 71  & 30 & 418  & 88.8\% & 92.7\% & 84.3\% & 88.3\%  \\
                         &                          &                                & MR-3 & 378 & 67  & 27 & 428  & 89.6\% & 93.3\% & 84.9\% & 88.9\%  \\
\cline{3-12}
                         &                          & \multirow{3}{*}{VinVL}         & MR-1 & 208 & 47  & 24 & 621  & 92.1\% & 89.7\% & 81.6\% & 85.4\%  \\
                         &                          &                                & MR-2 & 267 & 27  & 23 & 583  & 94.4\% & 92.1\% & 90.8\% & 91.4\%  \\
                         &                          &                                & MR-3 & 242 & 23  & 19 & 616  & 95.3\% & 92.7\% & 91.3\% & 92.0\%  \\
\cline{3-12}
                         &                          & \multirow{3}{*}{OFA}           & MR-1 & 199 & 39  & 27 & 635  & 92.7\% & 88.1\% & 83.6\% & 85.8\%  \\
                         &                          &                                & MR-2 & 200 & 17  & 27 & 656  & 95.1\% & 88.1\% & 92.2\% & 90.1\%  \\
                         &                          &                                & MR-3 & 206 & 26  & 21 & 647  & 94.8\% & 90.7\% & 88.8\% & 89.8\%  \\
\cline{3-12}
                         &                          & \multirow{3}{*}{Azure}         & MR-1 & 250 & 27 & 21 & 602 & 94.7\% & 92.3\% & 90.3\% & 91.2\% \\
                         &                          &                                & MR-2 & 312 & 22 & 29 & 537 & 94.3\% & 91.5\% & 93.4\% & 92.4\% \\
                         &                          &                                & MR-3 & 308 & 13 & 19 & 560 & 96.4\% & 94.2\% & 96.0\% & 95.1\% \\
\Xhline{0.8px}
\label{tab:RQ1}
\end{tabular}
\end{table*}

\begin{figure*}[h!]
    \centering
    \includegraphics[width=0.7\textwidth]{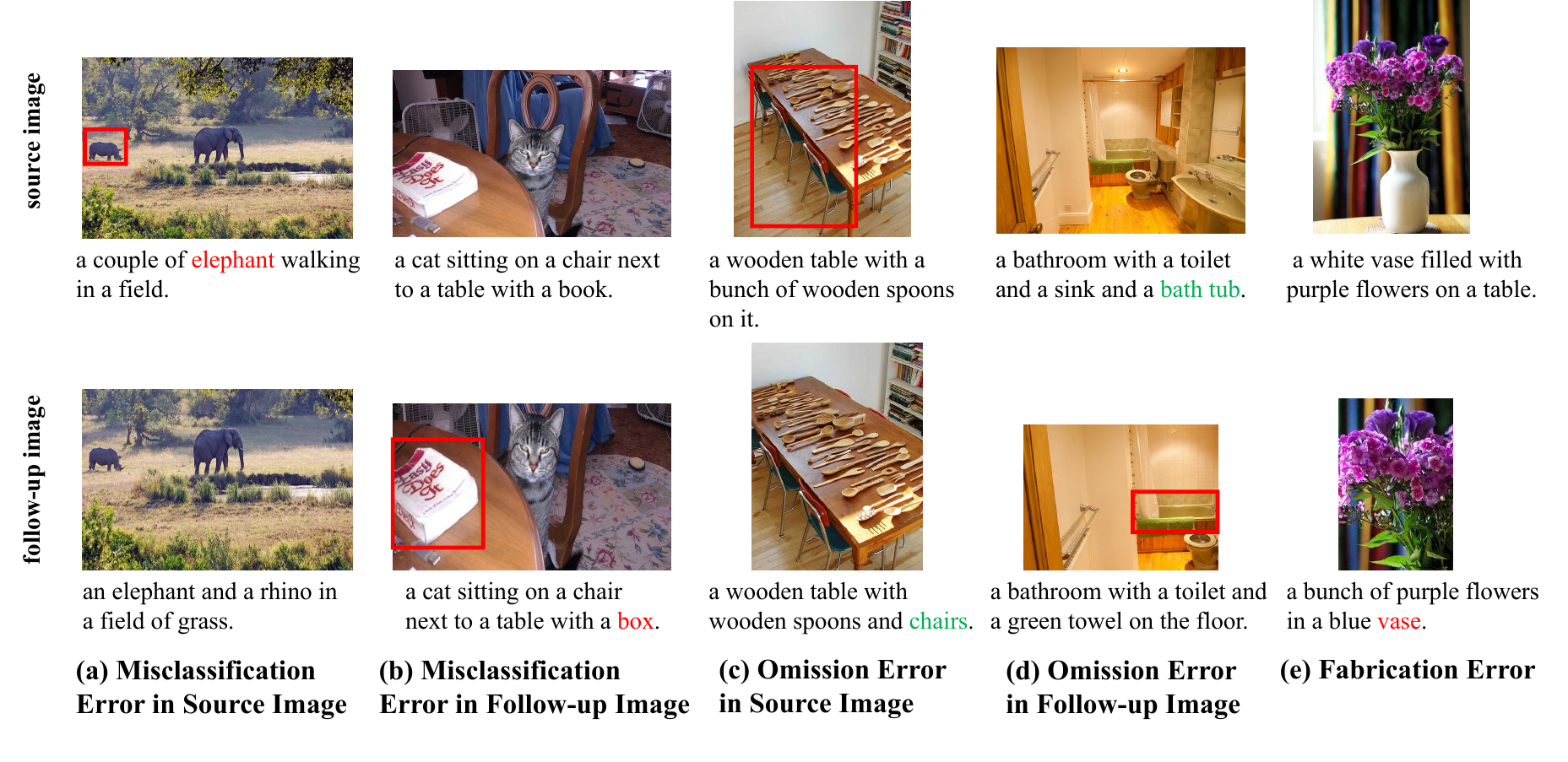}
    \caption{Examples of each type of erroneous captions reported by \methodname{}}
    \label{fig:errorExample}
\end{figure*}

In this RQ, we aim to evaluate the effectiveness of our approach by verifying whether the violations reported by \methodname{} indeed represent \textbf{actual violations}, i.e., TPs. 
We calculated the four metrics (Accuracy, Precision, Recall, and F1-score) of \methodname{}. In this RQ, we consider not only the precision of reported violations following the existing works but also the recall rate of our method. 
The recall rate can reflect the extent to which our method \textbf{covers} the errors in all MGs during the manual inspection process.

To thoroughly examine the effectiveness of \methodname{}, we report the results on two levels following~\cite{issre22MT4MOT}: \textbf{object-level} and \textbf{case-level}. An \textit{object-level violation} corresponds to an object in the caption that violates one of the rules defined in Section~\ref{subsec:violationCheck}. A \textit{case-level violation} will be reported if any object within a test case violates a rule. The object-level results provide a more detailed perspective. Each violation corresponds to an object within a test case. A case-level violation can encompass multiple objects. Table~\ref{tab:RQ1} shows the results of \methodname{} on different datasets, both at the object-level and case-level.

\subsubsection{Overall Performance} As shown in Table~\ref{tab:RQ1}, \methodname{} is able to detect quite a number of violations. For the 300 source test images in MSCOCO and PASCAL datasets, \methodname{} can find 218 to 726 violations at object-level successfully by adopting the predefined MRs. We can find that across different IC systems, \methodname{} can detect remarkable violations, showcasing the universality of our approach. 

In terms of precision, we found that the precision of the detected violations is mostly 80\% to 95\%. It demonstrates that most of the detected violations reflect actual violations. Given the fact that \textbf{our method requires no annotated information for the images}, the experimental results demonstrate the effectiveness of our approach. As for the recall, \methodname{} is also able to find around 70\% to 90\% of the errors in total. In conclusion, our approach is able to uncover a number of violations. The F1-score of our method is mostly 75\% to 90\%, suggesting its strong performance in terms of both precision and recall. These results confirm the overall effectiveness of our method.

\subsubsection{Types of Detected Errors} We examined the actual violations reported by \methodname{} and categorized them into three major types. It demonstrates the capability of our method to uncover various types of errors.

\textbf{Type 1: \textit{Misclassification Error.}} The SUT fails to classify a particular object in the test image correctly. Such type of error can further be divided into misclassification errors in source cases and misclassification errors in follow-up cases.
\begin{itemize}
    \item Type 1.1: \textit{Misclassification Error on Source Image.} As shown in Fig.~\ref{fig:errorExample}(a), the SUT misclassifies the \textit{``rhino''} in the source image as an \textit{``elephant''}.
    \item Type 1.2: \textit{Misclassification Error on Follow-up Image.} Similarly, as seen in Fig.~\ref{fig:errorExample}(b), the SUT mistakenly classifies the \textit{``book''} in the follow-up image as a \textit{``box''}.
\end{itemize}

\textbf{Type 2: \textit{Omission Error.}} The SUT fails to describe a salient object in the image. Such types of errors can also be categorized into omission errors on the source and follow-up images, respectively.

\begin{itemize}
    \item Type 2.1: \textit{Omission Error on Source Image.} If the source image and the follow-up image are deemed equivalent (as defined in Section ~\ref{subsec:violationCheck}) but the source output does not describe a salient object present in the follow-up output, it indicates an omission error in the source image. For example, in Fig.~\ref{fig:errorExample}(c), the output of the source image omits the \textit{``chair''} which is described in the output of the equivalent follow-up image.
    
    \item Type 2.2: \textit{Omission Error on Follow-up Image.} If a salient object described in the source caption does not disappear after transformation and is not further described in the follow-up caption, it implies an omission error in the follow-up image. For example, in Fig.~\ref{fig:errorExample}(d), \textit{``bath tub''} does not disappear in the follow-up image, yet the output of the follow-up image omits its description.
\end{itemize}

\textbf{Type 3: \textit{Fabrication Error.}} An object in the follow-up image completely disappears after the transformation, but it remains described in the follow-up output. This suggests that the SUT may \textbf{fabricate} certain objects based on the context of the current image. As shown in Fig.~\ref{fig:errorExample}(e), \textit{``vase''} disappears in the follow-up image, but the IC system fabricates its existence, maybe based on the presence of ``\textit{flower}''. Note that this type of error is not reported by existing works \cite{MetaIC_Pinjia, ROME_Pinjia}.

\begin{RQbox}{RQ1}
    {\methodname{} is able to utilize \textbf{unannotated} images to detect numerous errors across five IC systems, while existing works rely on pre-annotated labels. The experimental results show that \methodname{} achieves precision mostly in the range of 75\% to 90\%, which is comparable to existing works, but in the setting where no annotated information is given. Moreover, our method detects several types of errors, one of which has not been reported by existing works~\cite{MetaIC_Pinjia, ROME_Pinjia}.}
\end{RQbox}

\subsection{Eligibility of Test Cases (RQ2)}
\label{subsec:RQ2}

\begin{table*}[h!]
    \centering
    \scriptsize
    \caption{Comparison of Eligibility and Number of Actual Violations}
    \begin{tabular}{cccccc}
        \Xhline{0.8px}
        \textbf{Dataset} & \textbf{SUT} & \textbf{MR} & \textbf{\# Eligible Cases} & \textbf{\# Follow-up Images} & \textbf{\# Actual Violations (case-level)} \\
        \hline
        \multirow{10}{*}{MSCOCO}  & \multirow{5}{*}{Show and Tell} & MR-1 & 300 / 300 & 900 & 292 \\
        ~ & ~ & MR-2 & 300 / 300 & 900 & 320 \\
        ~ & ~ & MR-3 & 300 / 300 & 900 & 302 \\
        \cline{3-6}
        ~ & ~ & MetaIC & 46 / 300 & 121 & 116 \\
        ~ & ~ & ROME & 256 / 300 & 654 & {131} \\
        \cline{2-6}
        ~ & \multirow{5}{*}{VinVL} & MR-1 & 300 / 300 & 900 & 200 \\
        ~ & ~ & MR-2 & 300 / 300 & 900 & 277 \\
        ~ & ~ & MR-3 & 300 / 300 & 900 & 263 \\
        \cline{3-6}
        ~ & ~ & MetaIC & 46 / 300 & 121 & 58 \\
        ~ & ~ & ROME & 256 / 300 & 654 & 86 \\
        \hline
        \multirow{10}{*}{PASCAL}  & \multirow{5}{*}{Show and Tell} & MR-1 & 300 / 300 & 900 & 322 \\
        ~ & ~ & MR-2 & 300 / 300 & 900 & 362 \\
        ~ & ~ & MR-3 & 300 / 300 & 900 & 320 \\
        \cline{3-6}
        ~ & ~ & MetaIC & 60 / 300 & 154 & 128 \\
        ~ & ~ & ROME & 150 / 300 & 339 & 44 \\
        \cline{2-6}
        ~ & \multirow{5}{*}{VinVL} & MR-1 & 300 / 300 & 900 & 208 \\
        ~ & ~ & MR-2 & 300 / 300 & 900 & 267 \\
        ~ & ~ & MR-3 & 300 / 300 & 900 & 242 \\
       \cline{3-6}
        ~ & ~ & MetaIC & 60 / 300 & 154 & 49 \\
        ~ & ~ & ROME & 150 / 300 & 339 & 26 \\
        \Xhline{0.8px}
    \end{tabular}
    \label{tab:RQ2}
\end{table*}

This RQ aims to compare \methodname{} and existing methods in terms of the eligibility of source test cases. For this RQ, we selected two representative models, \textit{Show and Tell} and \textit{VinVL}, as the test objects. We can get similar conclusions with the other IC systems under our testing.

Both \methodname{} and existing methods can generate multiple images from a source image. We limited the number of follow-up images generated from a source image to ensure an equitable comparison. In this work, we generated three follow-up images for a source image. Employing scripts provided by the authors of MetaIC and ROME~\cite{RobustICCode}, we got the eligible cases and collected the violation detection results for these approaches. 

Table~\ref{tab:RQ2} shows the number of eligible test cases for different methods, as well as the actual violations revealed by these methods. Note that we selected the same 300 images from each dataset as the available test images.

In terms of the eligibility of test cases, \textbf{the existing approaches can use fewer images as the eligible source cases.} As shown in Table~\ref{tab:RQ2}, only 46 images from MSCOCO dataset and 60 images from PASCAL dataset can be used as the eligible source cases for MetaIC. Similarly, only 256 MSCOCO images and 150 PASCAL images can be used as eligible source cases for ROME. In contrast, our method does not impose requirements on the eligible source cases. It can use all the provided images to generate more follow-up images for testing. As a result, \textbf{\methodname{} can find a number of errors} ranging from 200 to 362, compared to 58 to 128 errors and 26 to 131 errors detected by MetaIC and ROME, respectively. This result shows that \methodname{} is effective in utilizing test images for testing, and is thereby able to uncover more errors in the IC systems.

We noticed that ROME can test 256 out of 300 images from the MSCOCO dataset. However, the number of eligible test cases decreases to 150 out of 300 on the test images from the PASCAL dataset. 
This discrepancy may be associated with the greater level of detail and comprehensiveness in the annotation of the MSCOCO dataset in contrast to the relatively fewer annotated objects in the PASCAL dataset, where object labels are also more scarce. This further underscores the dependency on annotated information of ROME.

\begin{RQbox}{RQ2}
    {{In general,} our approach achieves much higher eligibility on given test cases than MetaIC and ROME. The results suggest that \methodname{} can fully utilize source test cases and detect more violations than existing works in the provided images.}
\end{RQbox}

\subsection{Quality of Generated Test Cases (RQ3)}

\begin{table*}[h!]
    \scriptsize
    \centering
    \caption{Experimental Result of Effectiveness of Ambiguity Score in the Enhanced Follow-up Test Cases Generation Module}
    \begin{tabular}{ccccccccc}
        \Xhline{0.8px}
        \textbf{SUT} & \textbf{MR} & \textbf{Setting} & $N_{ambiguous}(\downarrow)$ & $N_{disappear}(\uparrow)$ & $N_{retain}(\uparrow)$ & \makecell{\textbf{\# Valid} \\ \textbf{\ \ Objects}} & \makecell{\textbf{\# Valid} \\ \textbf{\ \ Cases}} & \makecell{\textbf{\# Reported} \\ \textbf{Violations}} \\
        \Xhline{0.8px}
        \multirow{6}{*}{Show and Tell} & \multirow{2}{*}{MR-1} & Setting-1 & 1234 & 202 & 784 & 986 & 664 & 245 \\
        ~ & ~ & \methodname{} & \textbf{352} & 338 & 1530 & \textbf{1868} & \textbf{897} & 529 \\
        \cline{2-9}
        ~ & \multirow{2}{*}{MR-2} & Setting-1 & 1158 & 60 & 1002 & 1062 & 644 & 315 \\
        ~ & ~ & \methodname{} & \textbf{139} & 84 & 1997 & \textbf{2081} & \textbf{900} & 667 \\
        \cline{2-9}
        ~ & \multirow{2}{*}{MR-3} & Setting-1 & 1225 & 30 & 965 & 995 & 583 & 261 \\
        ~ & ~ & \methodname{} & \textbf{102} & 40 & 2078 & \textbf{2118} & \textbf{900} & 607 \\
        \cline{1-9}
        \multirow{6}{*}{VinVL} & \multirow{2}{*}{MR-1} & Setting-1 & 1518 & 226 & 755 & 981 & 653 & 165 \\
        ~ & ~ & \methodname{} & \textbf{519} & 400 & 1580 & \textbf{1980} & \textbf{900} & 371 \\
        \cline{2-9}
        ~ & \multirow{2}{*}{MR-2} & Setting-1 & 1306 & 74 & 1119 & 1193 & 688 & 233 \\
        ~ & ~ & \methodname{} & \textbf{180} & 105 & 2214 & \textbf{2319} & \textbf{900} & 541 \\
        \cline{2-9}
        ~ & \multirow{2}{*}{MR-3} & Setting-1 & 1362 & 33 & 1104 & 1137 & 618 & 224 \\
        ~ & ~ & \methodname{} & \textbf{141} & 37 & 2321 & \textbf{2358} & \textbf{900} & 502 \\
        \Xhline{0.8px}
    \end{tabular}
    \label{tab:RQ3.1}
\end{table*}

\begin{table*}[h!]
    \centering
    \scriptsize
    \caption{Experimental Result of Effectiveness of Diversity Score in the Enhanced Follow-up Test Cases Generation Module}
    \begin{tabular}{ccccc}
        \Xhline{0.8px}
        \textbf{SUT} & \textbf{MR} & \textbf{Setting} & \textbf{\# Coverage Distinct Objects}  & \textbf{\# Coverage Distinct Cases} \\
        \Xhline{0.8px}
        \multirow{6}{*}{Show and Tell} & \multirow{2}{*}{MR-1} & Setting-2 & 339 & 155 \\ 
        ~ & ~ & \methodname{} & \textbf{354(+4.42\%)} & \textbf{177(+14.19\%)} \\ 
        \cline{2-5}
        ~ & \multirow{2}{*}{MR-2} & Setting-2 & 399 & 169 \\ 
        ~ & ~ & \methodname{} & \textbf{439(+10.03\%)} & \textbf{183(+8.28\%)} \\ 
        \cline{2-5}
        ~ & \multirow{2}{*}{MR-3} & Setting-2 & 361 & 160 \\ 
        ~ & ~ & \methodname{} & \textbf{427(+18.28\%)} & \textbf{182(+13.75\%)} \\ 
        \cline{1-5}
        \multirow{6}{*}{VinVL} & \multirow{2}{*}{MR-1} & Setting-2 & 227 & 127 \\ 
        ~ & ~ & \methodname{} & \textbf{244(+7.49\%)} & \textbf{139(+9.45\%)} \\ 
        \cline{2-5}
        ~ & \multirow{2}{*}{MR-2} & Setting-2 & 307 & 144 \\ 
        ~ & ~ & \methodname{} & \textbf{365(+18.89\%)} & \textbf{169(+17.36\%)} \\ 
        \cline{2-5}
        ~ & \multirow{2}{*}{MR-3} & Setting-2 & 291 & 129 \\ 
        ~ & ~ & \methodname{} & \textbf{330(+13.40\%)} & \textbf{156(+20.93\%)} \\ 
        \Xhline{0.8px}
    \end{tabular}
    \label{tab:RQ3.2}
\end{table*}

To address the second and the third limitations, we designed the \textit{Enhanced Follow-up Test Cases Generation} module in Section~\ref{subsec:caseGeneration} to select images with lower ambiguity and higher diversity. This RQ aims to measure the quality of follow-up test cases generated by \methodname{}. Specifically, we validated the quality of our follow-up cases by comparing them with those generated with the settings that exclude $Score_{ambiguity}$ or $Score_{diversity}$ defined in Section~\ref{subsec:caseGeneration}. As stated in RQ2, we also considered two representative IC systems and we can draw the same conclusion on the other IC systems under our testing.

\textbf{Setting-1:} Setting-1 was set up to investigate the impact of $Score_{ambiguity}$. In this setting, the test cases were generated \textbf{without using $Score_{ambiguity}$} to select the candidate image set. That is, the first follow-up image was \textbf{randomly} selected from the set, and $Score_{diversity}$ was still considered for subsequent follow-up images. 

As mentioned in Section~\ref{subsec:caseGeneration}, we considered disappeared or retained objects as valid objects; and considered valid follow-up cases as the ones containing at least one suitable object for testing (i.e., disappeared object or retained object). In this RQ, we compared the number of valid objects, the number of valid follow-up cases, as well as the number of reported violations, delivered by \methodname{} with the ones delivered by Setting-1. The results are presented in Table~\ref{tab:RQ3.1}. 

According to the result shown in Table~\ref{tab:RQ3.1}, it is evident that $Score_{ambiguity}$ is very helpful in obtaining valid follow-up cases. The setting without considering this score generates a large portion of invalid follow-up cases. Note that \methodname{} may still generate some ambiguous objects, leading to a few invalid cases (3 out of 900). It is possible that all candidate images we generate contain a certain number of ambiguous objects. It may arise from the complexity of object positioning in the images, and it is possible that all candidate images we generate contain a certain quantity of ambiguous objects. Furthermore, the results show that \methodname{} can report \textbf{more violations}, which indicates that the design of $Score_{ambiguity}$ indeed helps to enhance the fault-detection capability.

\textbf{Setting-2:} The goal of $Score_{diversity}$ was to generate higher diverse test cases, allowing our method to assess the SUT's behavior on a wider range of objects and cases. In Section~\ref{subsec:caseGeneration}, we discussed distinct violations at object-level. In this section, we further considered violations related to different source images as \textit{distinct} violations at \textit{case-level}. Thus, the evaluation metric for this experiment is the coverage of distinct objects and distinct cases by violations reported by \methodname{} across all MGs corresponding to the same source image. To achieve this, we first obtained all objects associated with the same source image using semantic similarity tools and observed the number of distinct objects covered by the testing method. 

To investigate the diversity of the follow-up cases produced by \methodname{}, we set up Setting-2, where only $Score_{ambiguity}$ was used in generating subsequent test cases \textbf{without considering $Score_{diversity}$}. The experimental results for Setting-2 are presented in Table~\ref{tab:RQ3.2}. The results show that \methodname{} covers a higher number of distinct objects compared to Setting-2. Additionally, \methodname{} covers a greater number of source images at case-level. These findings indicate that, by generating diverse test cases, the violation reported by \methodname{} can cover more distinct images and distinct objects, thereby detecting more distinct violations.

\begin{RQbox}{RQ3}
    {Our approach is able to generate higher quality test cases with lower ambiguity and higher diversity, hence detecting more violations and distinct violations.}
\end{RQbox}

\subsection{Ablation Study for Violation Measurement (RQ4)}

In this RQ, we are interested in the effectiveness of several components of our method in detecting violations in generated MGs after generating follow-up test cases.

\begin{table*}[h!]
    \centering
    \scriptsize
    \caption{Ablation Study Result of Violation Detection for \methodname{} (Object-level)}
    \label{tab:RQ4(obj)}
    \begin{threeparttable}
        \begin{tabular}{ccccccccccl}
        \Xhline{0.8px}
        \textbf{SUT} & \textbf{MR} & \textbf{Group} & \textbf{TP} & \textbf{FN} & \textbf{FP} & \textbf{TN} & \textbf{Precision} & \textbf{Recall} & \textbf{F1-score} & \textbf{$\Delta$} \\
        \Xhline{0.8px}
        \multirow{12}{*}{Show and Tell} & \multirow{4}{*}{MR-1} & \methodname{} & 473 & 160 & 56 & 2789 & 89.4\% & 74.7\% & 81.4\% & / \\
        ~ & ~ & Group-1 & 391 & 238 & 115 & 2706 & 77.3\% & 62.2\% & 68.9\% & -12.5\% \\
        ~ & ~ & Group-2 & 253 & 238 & 31 & 2477 & 89.1\% & 51.5\% & 65.3\% & -16.1\% \\
        ~ & ~ & Group-3 & 488 & 145 & 256 & 2589 & 65.6\% & 77.1\% & 70.9\% & -10.5\% \\
        \cline{2-11}
        ~ & \multirow{4}{*}{MR-2} & \methodname{} & 597 & 160 & 70 & 3214 & 89.5\% & 78.9\% & 83.8\% & / \\
        ~ & ~ & Group-1 & 529 & 206 & 78 & 3004 & 87.1\% & 72.0\% & 78.8\% & -5.0\% \\
        ~ & ~ & Group-2 & 323 & 242 & 40 & 2818 & 89.0\% & 57.2\% & 69.6\% & -14.2\% \\
        ~ & ~ & Group-3 & 638 & 119 & 352 & 2932 & 64.4\% & 84.3\% & 73.0\% & -10.8\% \\
        \cline{2-11}
        ~ & \multirow{4}{*}{MR-3} & \methodname{} & 561 & 163 & 46 & 3447 & 92.4\% & 77.5\% & 84.3\% & / \\
        ~ & ~ & Group-1 & 477 & 201 & 45 & 3273 & 91.4\% & 70.4\% & 79.5\% & -4.8\% \\
        ~ & ~ & Group-2 & 298 & 232 & 24 & 2955 & 92.5\% & 56.2\% & 70.0\% & -14.3\% \\
        ~ & ~ & Group-3 & 607 & 117 & 263 & 3230 & 69.8\% & 83.8\% & 76.2\% & -8.1\% \\
        \hline
        \multirow{12}{*}{Oscar} & \multirow{4}{*}{MR-1} & \methodname{} & 398 & 142 & 57 & 3127 & 87.5\% & 73.7\% & 80.0\% & / \\
        ~ & ~ & Group-1 & 324 & 200 & 116 & 3012 & 73.6\% & 61.8\% & 67.2\% & -12.8\% \\
        ~ & ~ & Group-2 & 222 & 195 & 34 & 2848 & 86.7\% & 53.2\% & 66.0\% & -14.0\% \\
        ~ & ~ & Group-3 & 428 & 112 & 224 & 2960 & 65.6\% & 79.3\% & 71.8\% & -8.2\% \\
        \cline{2-11}
        ~ & \multirow{4}{*}{MR-2} & \methodname{} & 577 & 149 & 77 & 3555 & 88.2\% & 79.5\% & 83.6\% & / \\
        ~ & ~ & Group-1 & 469 & 197 & 96 & 3375 & 83.0\% & 70.4\% & 76.2\% & -7.4\% \\
        ~ & ~ & Group-2 & 335 & 228 & 48 & 3221 & 87.5\% & 59.5\% & 70.8\% & -12.8\% \\
        ~ & ~ & Group-3 & 656 & 70 & 354 & 3278 & 65.0\% & 90.4\% & 75.6\% & -8.0\% \\
        \cline{2-11}
        ~ & \multirow{4}{*}{MR-3} & \methodname{} & 593 & 119 & 68 & 3807 & 89.7\% & 83.3\% & 86.4\% & / \\
        ~ & ~ & Group-1 & 500 & 163 & 66 & 3609 & 88.3\% & 75.4\% & 81.4\% & -5.0\% \\
        ~ & ~ & Group-2 & 347 & 211 & 42 & 3366 & 89.2\% & 62.2\% & 73.3\% & -13.1\% \\
        ~ & ~ & Group-3 & 661 & 51 & 303 & 3572 & 68.6\% & 92.8\% & 78.9\% & -7.5\% \\
        \hline
        \multirow{12}{*}{VinVL} & \multirow{4}{*}{MR-1} & \methodname{} & 304 & 57 & 67 & 3286 & 81.9\% & 84.2\% & 83.1\% & / \\
        ~ & ~ & Group-1 & 226 & 118 & 129 & 3179 & 63.7\% & 65.7\% & 64.7\% & -18.4\% \\
        ~ & ~ & Group-2 & 190 & 115 & 45 & 3053 & 80.9\% & 62.3\% & 70.4\% & -12.7\% \\
        ~ & ~ & Group-3 & 301 & 60 & 265 & 3088 & 53.2\% & 83.4\% & 64.9\% & -18.1\% \\
        \cline{2-11}
        ~ & \multirow{4}{*}{MR-2} & \methodname{} & 474 & 51 & 67 & 3847 & 87.6\% & 90.3\% & 88.9\% & / \\
        ~ & ~ & Group-1 & 393 & 97 & 85 & 3599 & 82.2\% & 80.2\% & 81.2\% & -7.7\% \\
        ~ & ~ & Group-2 & 306 & 133 & 44 & 3522 & 87.4\% & 69.7\% & 77.6\% & -11.4\% \\
        ~ & ~ & Group-3 & 497 & 28 & 363 & 3551 & 57.8\% & 94.7\% & 71.8\% & -17.2\% \\
        \cline{2-11}
        ~ & \multirow{4}{*}{MR-3} & \methodname{} & 452 & 77 & 50 & 4101 & 90.0\% & 85.4\% & 87.7\% & / \\
        ~ & ~ & Group-1 & 377 & 107 & 61 & 3840 & 86.1\% & 77.9\% & 81.8\% & -5.9\% \\
        ~ & ~ & Group-2 & 301 & 140 & 34 & 3688 & 89.9\% & 68.3\% & 77.6\% & -10.1\% \\
        ~ & ~ & Group-3 & 486 & 43 & 363 & 3788 & 57.2\% & 91.9\% & 70.5\% & -17.1\% \\
        \hline
        \multirow{12}{*}{OFA} & \multirow{4}{*}{MR-1} & \methodname{} & 218 & 87 & 70 & 3379 & 75.7\% & 71.5\% & 73.5\% & / \\
        ~ & ~ & Group-1 & 141 & 124 & 49 & 2981 & 74.2\% & 53.2\% & 62.0\% & -11.5\% \\
        ~ & ~ & Group-2 & 152 & 139 & 125 & 3282 & 54.9\% & 52.2\% & 53.5\% & -20.0\% \\
        ~ & ~ & Group-3 & 229 & 76 & 221 & 3228 & 50.9\% & 75.1\% & 60.7\% & -12.9\% \\
        \cline{2-11}
        ~ & \multirow{4}{*}{MR-2} & \methodname{} & 306 & 59 & 80 & 4077 & 79.3\% & 83.8\% & 81.5\% & / \\
        ~ & ~ & Group-1 & 244 & 88 & 78 & 3846 & 75.8\% & 73.5\% & 74.6\% & -6.9\% \\
        ~ & ~ & Group-2 & 173 & 120 & 51 & 3512 & 77.2\% & 59.0\% & 66.9\% & -14.6\% \\
        ~ & ~ & Group-3 & 338 & 27 & 334 & 3823 & 50.3\% & 92.6\% & 65.2\% & -16.3\% \\
        \cline{2-11}
        ~ & \multirow{4}{*}{MR-3} & \methodname{} & 295 & 67 & 54 & 4344 & 84.5\% & 81.5\% & 83.0\% & / \\
        ~ & ~ & Group-1 & 244 & 87 & 55 & 4080 & 81.6\% & 73.7\% & 77.5\% & -5.5\% \\
        ~ & ~ & Group-2 & 160 & 122 & 37 & 3668 & 81.2\% & 56.7\% & 66.8\% & -16.2\% \\
        ~ & ~ & Group-3 & 330 & 32 & 292 & 4106 & 53.1\% & 91.2\% & 67.1\% & -15.9\% \\
        \Xhline{0.8px}
    \end{tabular}
    \begin{tablenotes}
        \footnotesize
        \item[*] Group-1: \methodname{} w/o \textit{object detection} component
        \item[*] Group-2: \methodname{} w/o \textit{occlusion-based localization} component
        \item[*] Group-3: \methodname{} w/o \textit{semantic similarity} component
    \end{tablenotes}
    \end{threeparttable}
\end{table*}

We designed three groups of method variants to examine the impact of the \textit{object detection}, the \textit{occlusion-based localization}, and the \textit{semantic similarity} components on \methodname{} respectively. Object detection and occlusion-based localization constitute the key components of the \textit{Visual-Caption Object Alignment} module. Therefore, we separately designed ablation groups for these two components. In addition, we employed FastText~\cite{fasttext} and WordNet tools~\cite{wordnet} to assess the semantic similarity between two nouns to determine whether two objects belong to the same category. As such, we designed ablation experiments specifically targeting word semantic similarity as well.

\textbf{Group-1: Ablation of the Object Detection Component}. In this group, we solely employed occlusion-based localization to align objects between images and captions. As the occlusion-based localization individually locates objects in captions, it can function as an independent localization approach. 

\textbf{Group-2: Ablation of the Occlusion-based Localization Component}: To ablate the occlusion-based localization component, we only utilized OD to locate in this group. It is worth noting that OD may omit some objects, meaning that some objects in the captions might not have corresponding labels in the results of OD models. We classified such objects as non-violations.

\textbf{Group-3: Ablation of the Semantic Similarity Component}: In Group-3, we did not employ the semantic similarity tools for word comparison. 
For the two nouns in the captions, we considered them to belong to the same object class only when the two nouns are \textbf{an exact literal match}.

The ablation results on MSCOCO are presented in Table~\ref{tab:RQ4(obj)}. The last column of the table shows the decrease in F1-score for the ablation groups compared to \methodname{}. F1-score combines the precision and recall of a method, and it is evident that \methodname{} achieves the highest F1-score compared to the three ablation groups. This demonstrates that all the ablated components have a positive impact on our approach.

Firstly, we considered the ablation study for the \textit{Visual-Caption Object Alignment} module. We noted that \textbf{the recall values for Group-2 are significantly lower} than those of \methodname{}. This discrepancy arises because OD does not comprehensively cover the objects in the caption. For example, OD cannot locate the object ``drink'' described in the caption of Fig.~\ref{fig:od}(a) due to the limitation of the semantic similarity tool. As a result, OD cannot detect the violations, causing more occurrences of false negatives and a lower recall value. The results indicate that the occlusion-based localization component effectively complements OD to align the objects between the caption and the image. Furthermore, for some MRs and IC systems, \methodname{} achieves significantly higher precision.

Regarding the semantic similarity component, we can observe that Group-3, while not experiencing a significant decrease in F1-score, has \textbf{significantly lower precision} compared to \methodname{}. This is understandable since when semantic similarity is not used, some nouns that should belong to the same class (e.g., \textit{``person''} and \textit{``man''}) may be mistakenly considered different object categories, resulting in more false positives.

\begin{RQbox}{RQ4}
    {\methodname{} achieves the highest F1-score compared with these three ablation groups, which indicates that these three components yield positive effects and contribute to the enhancement of the performance of our approach.}
\end{RQbox}

\subsection{Comparison of Different Models, Datasets, and Transformation Methods (RQ5)}
\label{subsec:RQ5}

In this RQ, we started with an overall comparison of the number of actual violations detected by our \methodname{} in different models and datasets. Then, we categorized these errors into their respective error types and performed a more detailed comparison by exploring the number of each error type found in each scenario.

\begin{figure*} [h]
	\centering
	\subfloat[Comparison result on the {MSCOCO} dataset]{
		\includegraphics[width=2.7 in]{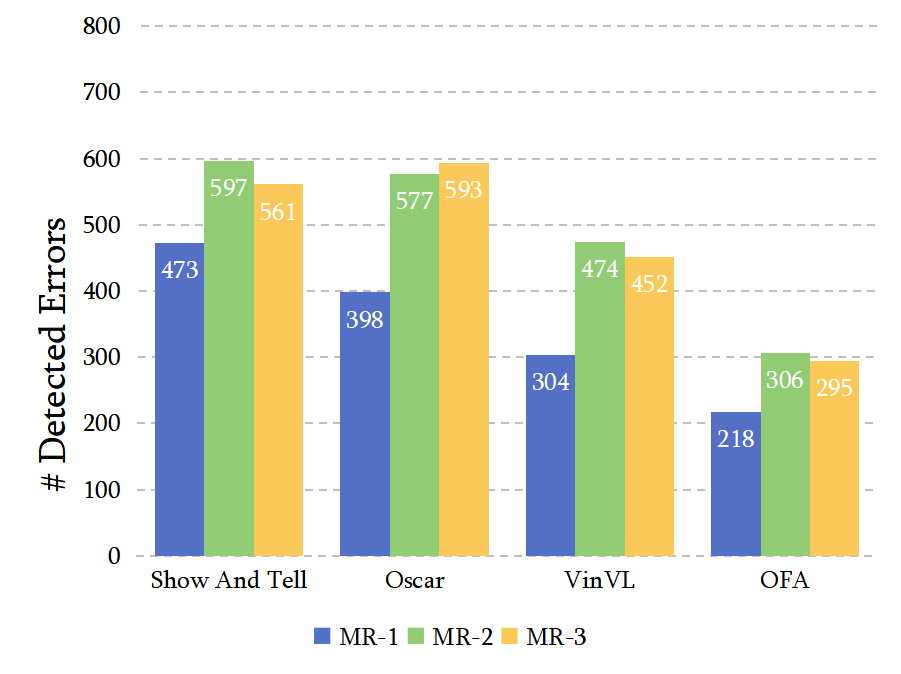}}
	\subfloat[Comparison result on the {PASCAL} dataset]{
		\includegraphics[width=2.7 in]{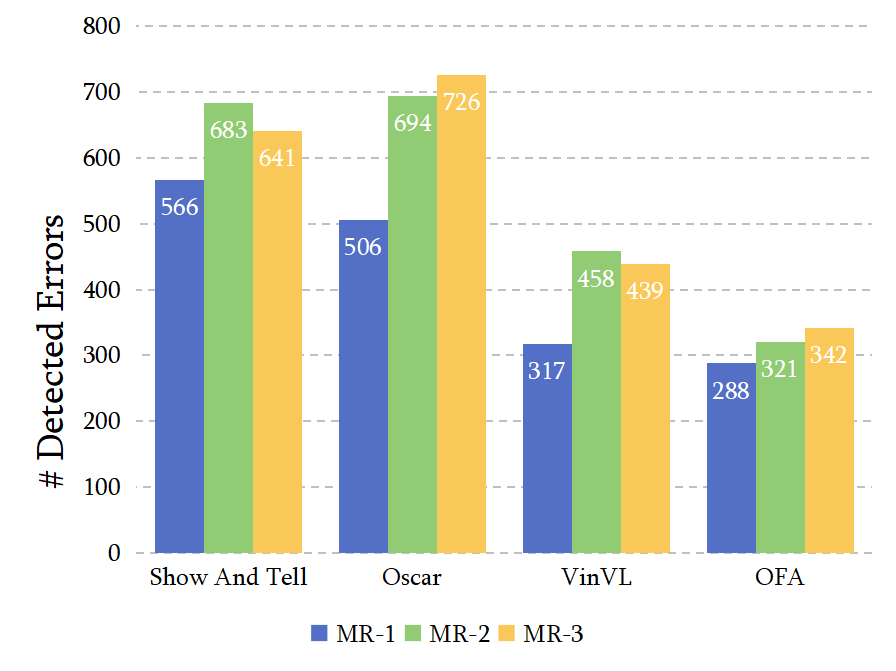}}
	\caption{Comparison of different models and MRs on number of detected errors}
	\label{fig:RQ5}
\end{figure*}

\begin{table*}[h!]
    \renewcommand{\arraystretch}{1.2}
    \centering
    \scriptsize
    \caption{Comparison of Error Number for Each Type in Different Models and MRs}
    \label{tab:RQ5}
    \begin{tabular}{l|ccc|ccc|ccc|ccc}
        \Xhline{0.08em}
        \textbf{} & \multicolumn{3}{c|}{\textbf{Show and Tell}} & \multicolumn{3}{c|}{\textbf{Oscar}} & \multicolumn{3}{c|}{\textbf{VinVL}} & \multicolumn{3}{c}{\textbf{OFA}} \\
        \hline
        \hline
        \textbf{Error Type} & MR-1 & MR-2 & MR-3 & MR-1 & MR-2 & MR-3 & MR-1 & MR-2 & MR-3 & MR-1 & MR-2 & MR-3 \\ \hline
        \textbf{Type 1.1} & 57 & 60 & 83 & 17 & 31 & 36 & 22 & 34 & 21 & 15 & 15 & 16 \\
        \textbf{Type 1.2} & 60 & 77 & 92 & 59 & 94 & 107 & 16 & 32 & 27 & 19 & 32 & 25 \\
        \textbf{Type 2.1} & 129 & 194 & 175 & 121 & 187 & 196 & 103 & 183 & 188 & 61 & 111 & 112 \\
        \textbf{Type 2.2} & 178 & 248 & 202 & 173 & 256 & 246 & 139 & 215 & 211 & 85 & 135 & 139 \\
        \textbf{Type 3} & 49 & 18 & 6 & 28 & 9 & 8 & 24 & 10 & 5 & 39 & 15 & 3 \\
        \Xhline{0.8px}
    \end{tabular}
\end{table*}

Fig.~\ref{fig:RQ5} presents the number of actual violations detected by \methodname{} on four different models using the three input transformation methods mentioned in this paper, separately for the MSCOCO and PASCAL datasets.

Firstly, concerning different models, we found that \methodname{} have detected fewer errors in most cases for models that reported better performance. This suggests that these models exhibit higher robustness when subjected to our testing method, which aligns with intuition.
Secondly, we observed that \methodname{} can consistently detect more errors in test images originating from the PASCAL dataset compared to those from the MSCOCO dataset. This may be due to the fact that existing models are primarily trained and validated on the MSCOCO dataset. This finding highlights the potential benefits of using diverse sources of test data instead of relying solely on model-provided validation datasets to uncover more potential errors~\cite{fse21-MT4MRC}.

More specifically, Table~\ref{tab:RQ5} shows the number of errors for each type detected by \methodname{} using different input transformation methods for MR on four models on the MSCOCO dataset. Based on the result, we have the following interesting findings.

Through further exploration, we found that the number of misclassification errors (Type 1) decreased most noticeably as model performance improved. This suggests that as model performance improves, the probability of misclassifying objects in the images decreases significantly. However, for fabrication errors (Type 3), we did not observe a similar pattern. This type of error does not decrease significantly with improved model performance. This suggests that these higher-performing models may still \textit{fabricate} objects based on the context of the current image. This finding suggests the need for model developers to focus on designing models specifically to reduce the likelihood of such errors.

\begin{table*}[h!]
    \renewcommand{\arraystretch}{1.2}
    \centering
    \scriptsize
    \caption{Comparison of Error Number and Average Number of Objects in Different Models}
    \label{tab:obj}
    \begin{tabular}{c|cccc|cccc}
        \Xhline{0.08em}
        \multirow{2}{*}{}                      & \multicolumn{4}{c|}{MSCOCO}           & \multicolumn{4}{c}{PASCAL}           \\\cline{2-9}
        ~                          & Show and Tell & Oscar & VinVL & OFA  & Show and Tell & Oscar & VinVL & OFA  \\\hline \hline
        Average \#Objects in Caption & 2.47          & 2.71  & 2.78  & 2.79 & 2.32          & 2.50  & 2.47  & 2.53 \\\hline
        \#Error in MR-1              & 473           & 398   & 304   & 218  & 566           & 506   & 317   & 288  \\
        \#Error in MR-2              & 597           & 577   & 474   & 306  & 683           & 694   & 458   & 321  \\
        \#Error in MR-3              & 561           & 593   & 452   & 295  & 641           & 726   & 439   & 342  \\\Xhline{0.08em}
    \end{tabular}
\end{table*}

Regarding different MRs with distinct transformation methods, we find that MR-1 (cropping) can detect more fabrication errors, while fewer errors of the other two types. Compared to stretching (which can only be applied along one axis) and rotation (which is limited to some fixed angles), cropping images is a more flexible approach. This suggests that applying more flexible transformation methods may help detect more \textit{fabrication errors}.

Furthermore, to explore the relationship between the number of objects in a caption and the number of detected errors at \textit{object-level}, we calculated the average number of objects in 300
source captions for each model, and compared the number of detected errors (TP) among the testing results of three MRs. The results are shown in Table~\ref{tab:obj}. According to the results, we did not observe any obvious relationship between the number of objects in a caption and the number of detected TPs. For example, for the results in the MSCOCO dataset, \textit{OFA} generates more objects in the output captions than \textit{Show and Tell}, but \textit{OFA} reveals fewer TPs in the testing results of all MRs. However, we can also find that \textit{Oscar} generates more objects in captions than \textit{Show and Tell}, but \textit{Oscar} reveals more TPs in the testing result of MR-3.

\begin{RQbox}{RQ5}
    {We found that \methodname{} can detect more violations using images from the PASCAL dataset compared to MSCOCO (which is used for training). This indicates that testers can benefit from using test cases from various datasets. Moreover, we found that current IC systems still suffer from ``fabricating'' objects that do not exist in images.}
\end{RQbox}

\section{Discussion}
\label{sec:discuss}
\subsection{Realism of Generated Test Images}
\label{subsec:discussreality}

As mentioned in~\cite{NaturalOD, AEON}, only considering the number of the reported violations is not enough. More importantly, we should reveal more meaningful violations.
Obviously, generating unrealistic follow-up test cases will lead to less meaningful violations. Thus, we should pay attention to the realism of the generated follow-up test cases to better simulate real-world scenarios. As discussed in Section~\ref{subsec:motivation}, existing approaches may still generate unrealistic images. 

\begin{figure}[h!]
    \centering
    \includegraphics[width=0.48\textwidth]{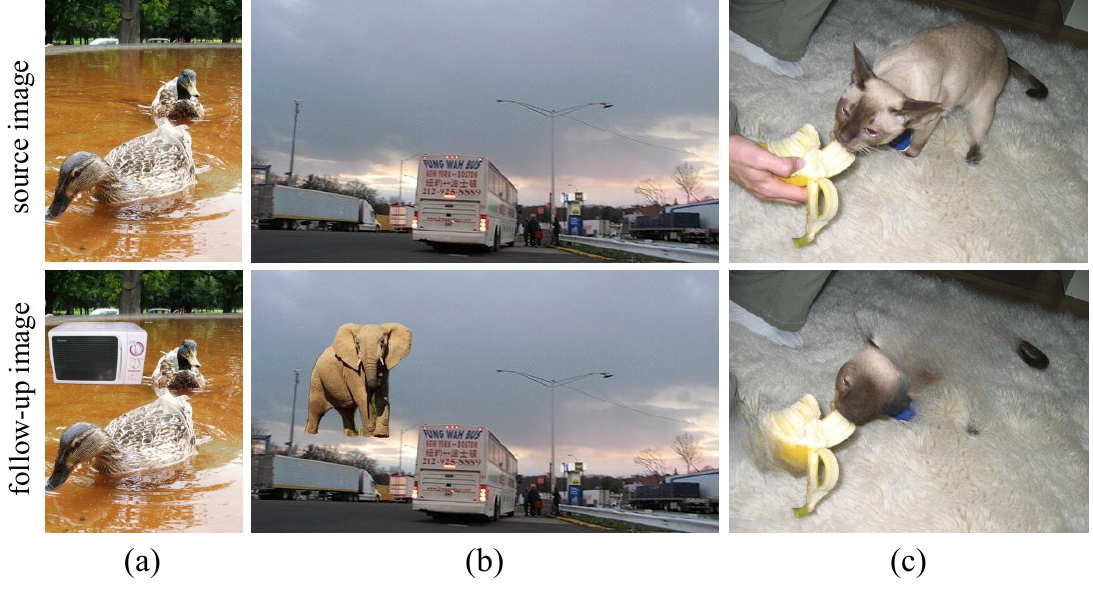}
    \caption{Unrealistic cases generated by existing methods}
    \label{fig:unrealCase}
\end{figure}

We found the following scenarios in which existing methods generate unrealistic images. Regarding MetaIC, we observed that (1) the inserted objects may not match the context of the background image, such as inserting an \textit{oven} into the pond in Fig.~\ref{fig:unrealCase}(a); (2) the size of the inserted object may not correspond to the natural proportions relative to the other objects in the background image, such as the tiny \textit{sofa} in Fig.~\ref{fig:existmethods}(b); and (3) the inserted object may lack adequate support, such as inserting an \textit{elephant} into the sky in Fig.~\ref{fig:unrealCase}(b). In fact, these scenarios do not exist in isolation. Many images generated by MetaIC suffer from multiple issues that impact their realism.

Regarding ROME, we observed that (1) ROME may result in unnatural shadows or reflections for objects after their melting. For example, after melting the oranges in Fig.~\ref{fig:existmethods}(c), ROME did not correctly delete the corresponding shadows~\cite{reality}; and (2) ROME may only \textbf{partially} melt objects, potentially resulting in absurd follow-up images. As illustrated in Fig.~\ref{fig:unrealCase}(c), ROME does not completely eliminate the \textit{cat}, instead, it melts the body of the \textit{cat} while preserving its head.

MetaIC and ROME do not completely avoid generating unrealistic images. In comparison, our approach can generate more realistic follow-up images, since it uses the reduction-based transformations that do not introduce any additional content into the image.

\subsection{False Positives}
To further understand the testing results of \methodname{}, we compiled a summary outlining the typical causes of FPs.

\begin{figure}[h!]
    \centering
    \includegraphics[width=0.4\textwidth]{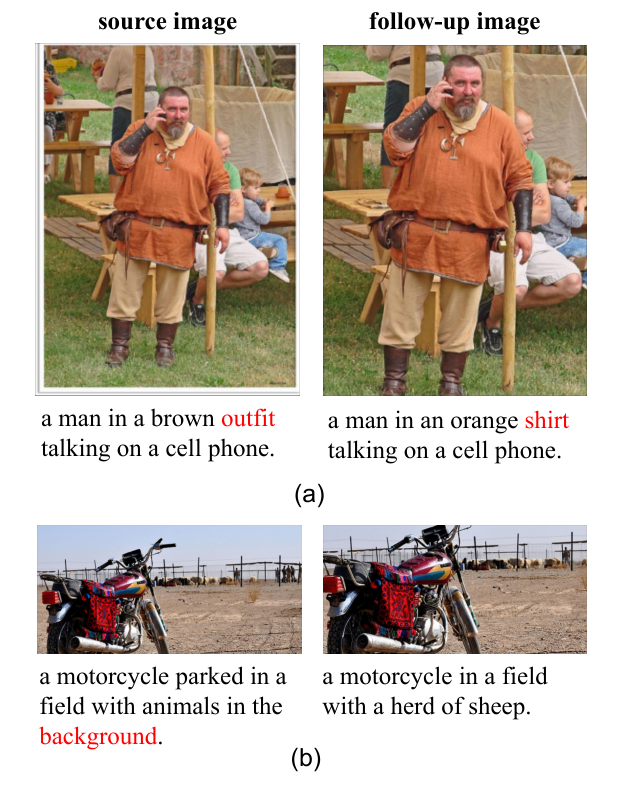}
    \caption{False positives examples generated by \methodname{}}
    \label{fig:falsepositive}
\end{figure}

First, FPs may be caused by the limitations inherent in \textit{semantic similarity} tools. Although \methodname{} employed semantic similarity tools in Section~\ref{subsec:loc} to verify whether two objects belong to the same category, misclassification, and subsequent false positives may still occur. For example, the tool categorizes ``outfit'' and ``shirt'' into different objects in Fig.~\ref{fig:falsepositive}(a), and reports a violation incorrectly. Furthermore, the integration of semantic similarity tools within the \textit{Visual-Caption Object Alignment} module may inadvertently compromise object localization results, thereby contributing to false positives.

Second, FPs may arise due to \textit{abstract nouns}. Although \methodname{} adopts certain rules to filter out specific nouns, such as removing nouns before ``of'', it still fails to filter out all abstract nouns, which often lack tangible counterparts. For example, as shown in Fig.~\ref{fig:falsepositive}(b), the absence of ``background'' in the follow-up caption can also cause an FP.

\subsection{Efficiency of Testing Process}
In this section, we compared the efficiency of \methodname{} and the existing IC testing approaches. The total testing time cost of \methodname{} can be divided into three parts: the time costs for \textit{localization}, \textit{follow-up images generation}, and \textit{violation measurement}, corresponding to the three components of \methodname{}. To assess the efficiency of \methodname{} and existing approaches, we ran each method five times individually and recorded the average time cost from generating a follow-up image to checking violations based on the follow-up image. Table~\ref{tab:time} shows the average time cost of each MR in \methodname{}, MetaIC, and ROME. The data presented in the table is based on the OFA model. We found that the other models under our testing exhibited similar trends.

\begin{table*}[h!]
    \renewcommand{\arraystretch}{1.2}
    \centering
    \scriptsize
    \caption{Comparison of Average Time Cost in Testing Process}
    \label{tab:time}
    \begin{tabular}{c|c|c|c|c}
        \Xhline{0.08em}
        \multirow{2}{*}{} & \multicolumn{4}{c}{\textbf{Average Time Cost}} \\
        \cline{2-5}
        & \textbf{Localization} & \textbf{Follow-up Images Generation} & \textbf{Violation Measurement} & \textbf{Total} \\
        \hline \hline
        MR-1 & \multirow{3}{*}{2.62s} & 0.45s & 0.172s & 3.242s \\
        MR-2 & ~ & 0.68s & 0.169s & 3.469s \\
        MR-3 & ~ & 0.71s & 0.167s & 3.497s \\
        \hline
        MetaIC & - & 3.29s & 0.144s & 3.434s \\
        ROME   & - & 0.69s & 0.166s & 0.856s \\
        \Xhline{0.08em}
    \end{tabular}
\end{table*}

As illustrated in Table~\ref{tab:time}, the total time cost of \methodname{} with different MRs is comparable, and the total time cost of \methodname{} is relatively higher than the total time cost of ROME. However, since \textbf{both MetaIC and ROME rely on pre-annotated information} about the position or corresponding masks of the objects in the image, they do not need the time for locating the objects. However, the time cost in localization process is the key factor of the higher total time cost of \methodname{} in comparison to ROME. As detailed in Section~\ref{subsec:motivation}, one of the primary motivations of our approach is to eliminate the reliance on manually annotated information. We think that this trade-off in time is worthwhile. Furthermore, the extra time cost of our method for localization is 2.62s, which is also considerably less than the time cost of human annotation.

\subsection{Threats to Validity}

\textbf{The first threat to validity} relates to the representativeness of the object models and dataset in our evaluation. To counter this threat, we considered five typical IC systems following \cite{ROME_Pinjia}, including four IC models and a commercial API. Regarding the representativeness of the dataset, we extended our data sources by not only using the widely used MSCOCO dataset for IC but also incorporating another dataset PASCAL. While PASCAL may not be a conventional choice for IC tasks, it is widely employed for segmentation-related assessments. Its image content is from various real-world scenarios and is proper for describing. We contend that testing with data from diverse sources is essential for a more thorough evaluation.

\textbf{The second threat to validity} concerns the performance limitation of the POS tagging tools and semantic similarity tools. We empolyed POS tagging tools implemented by Stanza~\cite{stanza}, which may misclassify certain words, resulting in false positives. Furthermore, we utilized word vectors provided by FastText~\cite{fasttext} and calculated their similarity to assess word similarity, which could occasionally lead to errors. In the future, we will explore alternative tools to enhance the accuracy of our tool.

\textbf{The third threat to validity} is about the validity of the measurement result. To evaluate the performance of \methodname{}, we manually inspected the testing result to decide the errors and analyze the type of TPs.
There may exist inaccuracies in the manual identification. To address this threat, we recruited three graduate students who were proficient in English to finish the above tasks. Fleiss's Kappa~\cite{fleisskappa} is a measure for the reliability of agreement between different dependent raters, and it is considered as a generalization of Cohen's kappa for more than two raters. In our manual inspection process, the Fleiss's Kappa are 0.876 and 0.891 in the first step (identify errors) and the second step (analyze the type of errors) respectively, indicating a high level of agreement.

\section{Related Work}
\label{sec:RelatedWork}
Recently, AI-driven software systems have been widely used in daily life. The widespread adoption of intelligent software has raised concerns about its security, particularly in critical domains such as autonomous driving~\cite{driving} and healthcare~\cite{health}. Thus, testing has been demonstrated to be an important way to identify potential issues and enhance the robustness of AI-driven software~\cite{testSurvey, DeepTest, DeepXplore, DeepBackground, DLTesting, DLFuzz}. MT is thought as an effective testing technique to tackle the oracle problem, and researchers have proposed several approaches that adopt MT in testing crucial fields~\cite{LiRTest, MetaOD, MetaVQA, qaasker, ROME_Pinjia}. In this section, we will introduce several MT approaches along with their corresponding MRs.

\textbf{Metamorphic Testing for CV Software.} Computer Vision (CV) refers to tasks that automatically extract, analyze, and understand valuable information from images. To test CV software, researchers have explored many transformation methods to conduct MT. For autonomous driving tasks, Tian et al.~\cite{DeepTest} and Zhang et al.~\cite{DeepRoad} introduced several real-world transformations to stimulate different driving conditions like blurring, rain, and fog. For the image classification task,  Dwarakanath et al.~\cite{DwarakanathMT} introduced several image transformations including permuting image channels (i.e., RGB channels), permuting convolution operation order, normalizing images, and scaling images. Zhang et al.~\cite{DeepBackground} introduced transformations that change the background of the test image. For the OD task, Wang et al.~\cite{MetaOD} inserted an object instance into the background image to generate a synthetic image. Shao et al.~\cite{3DRestruction} generated synthetic images by adopting 3D reconstruction of objects in real images. For multiple object tracking tasks, Xie et al.~\cite{issre22MT4MOT} introduced several transformations to simulate real-world scenarios and detect meaningful violations. In this work, we apply a class of reduction-based affine transformations to transform test images. Different from these works, our work focused on testing IC systems. We have addressed several limitations of the existing IC testing methods.

\textbf{Metamorphic Testing for Multimodal Software.} Multimodal tasks refer to those that involve multiple modalities such as image, video, text, and audio~\cite{multimodalSurvey}. These tasks include IC, Audio Speech Recognition (ASR), Image-Text Retrieval (ITR), and Visual Question Answering (VQA). Yu et al. proposed MetaIC~\cite{MetaIC_Pinjia} and ROME~\cite{ROME_Pinjia} that adopted MT for testing IC systems. Ji et al. proposed ASRTest~\cite{ASRTest} that applied characteristics mutation, noise injection, and reverberation simulation on speeches to test ASR. Yuan et al. proposed MetaVQA~\cite{MetaVQA} which simultaneously transforms the input images and questions. MetaVQA transformed an image with its question to one or a set of sub-images and sub-questions to test VQA. 
Our work involves aligning the objects in visual and textual modalities to facilitate checking the changes in the described object after image transformation. This enables us to design effective MRs for testing multi-modal IC tasks.

\section{Conclusion and Future Work}
\label{sec:Conclusion}
In this paper, we propose a metamorphic testing method for IC systems, \methodname{}, to address the limitations of the existing methods. Specifically, \methodname{} uses a localization method that combines OD and an occlusion-based method to address the limitation of relying on pre-annotated information. Furthermore, by leveraging reduction-based transformations that do not artificially manipulate any content in the original images, our method effectively avoids generating unreal follow-up test cases. At the same time, our method imposes no restrictions on the transformation process, allowing us to utilize more images for testing and increasing the eligibility of provided images. To enhance the quality of generated follow-up images, we design a dynamic strategy to select images with lower ambiguity and higher diversity, thereby enhancing the fault-detection capability of \methodname{}. The experimental results indicate that \methodname{} can sufficiently leverage provided test images to generate follow-up cases of good realism, and effectively detect a great number of distinct violations, without the need for any pre-annotated information.

In the future, we plan to implement more reduction-based transformations (such as shear and zoom) and embed them into our approach. This will enlarge the properties of IC systems that our method can examine. It also helps to validate the adaptability of our approach as a framework across different transformation methods. Additionally, we will test more IC software using our method, especially the large language models. We will also try various POS tools to find the most capable tools to enhance the performance of our approach.

\section*{Acknowledgments}
This work was supported by National Natural Science Foundation of China (Grant No. 62250610224). We sincerely appreciate the valuable suggestions from the anonymous reviewers for our paper.

\bibliographystyle{IEEEtran}
\bibliography{references}

\begin{IEEEbiography}[{\includegraphics[width=1in,height=1.25in,clip,keepaspectratio]{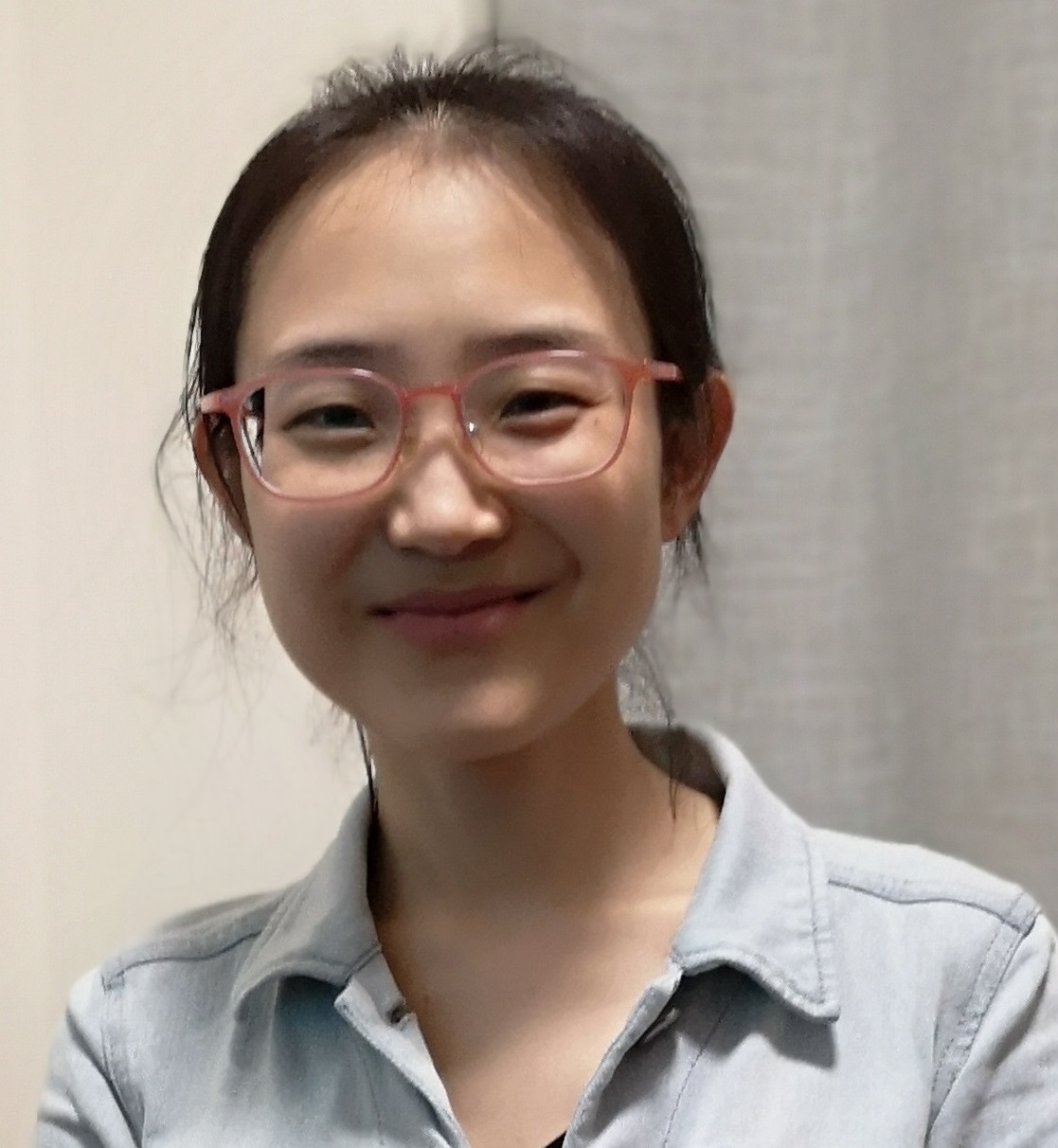}}]{Xiaoyuan Xie} received the BSc and MPhil degrees in computer science from Southeast University, China, in 2005 and 2007, respectively, and the PhD degree in computer science from the Swinburne University of Technology, Australia, in 2012. She is currently a professor in School of Computer Science, Wuhan University, China. Her research interests include software analysis, testing, debugging, and search-based software engineering. She is a member of the IEEE.
\end{IEEEbiography}

\begin{IEEEbiography}[{\includegraphics[width=1in,height=1.25in,clip,keepaspectratio]{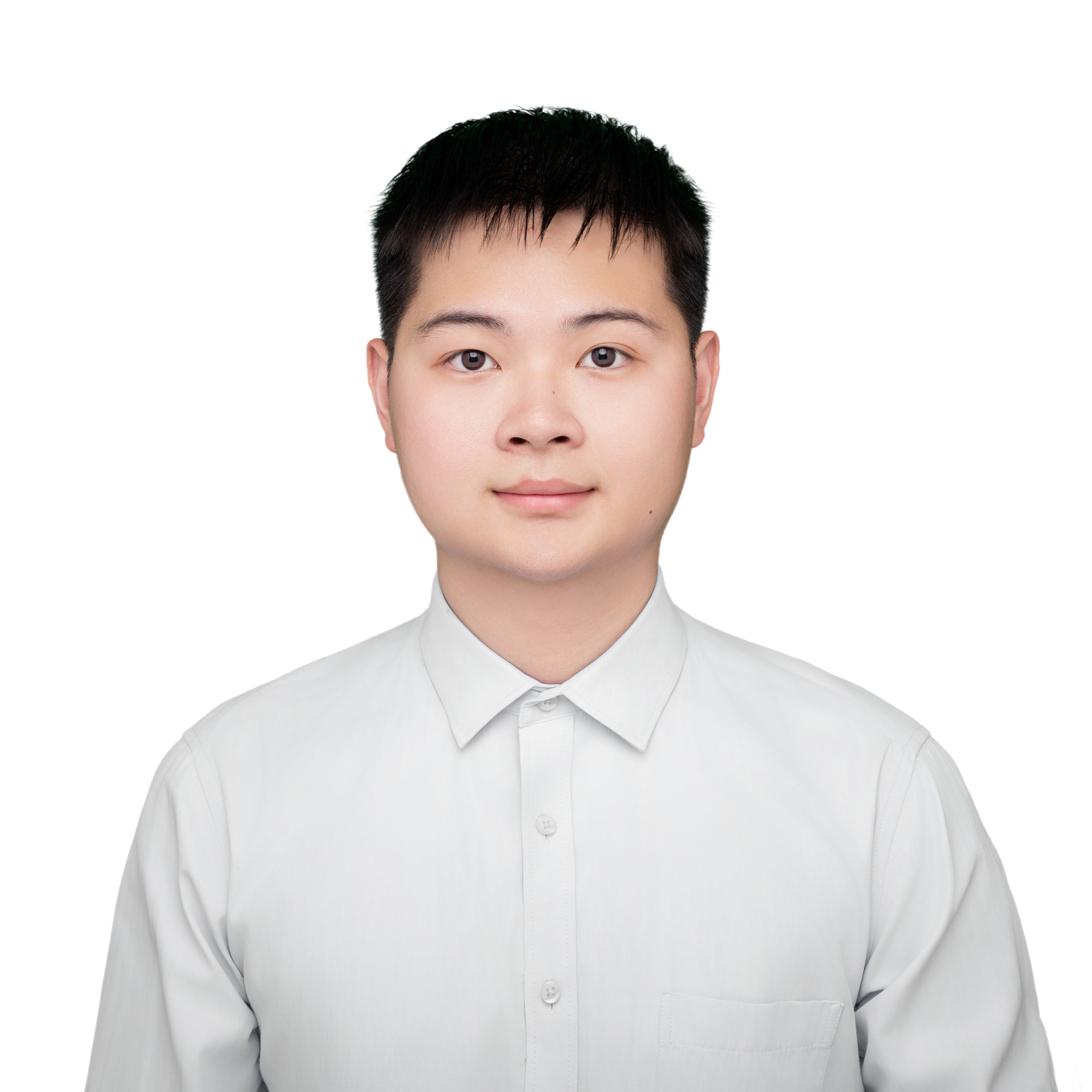}}]{Xingpneg Li} received his bachelor's degree in software engineering at Wuhan University, China, in 2022. He is currently working toward a master’s degree at the School of Computer Science, Wuhan University, China. His research interests include software testing and metamorphic testing.
\end{IEEEbiography}

\begin{IEEEbiography}[{\includegraphics[width=1in,height=1.25in,clip,keepaspectratio]{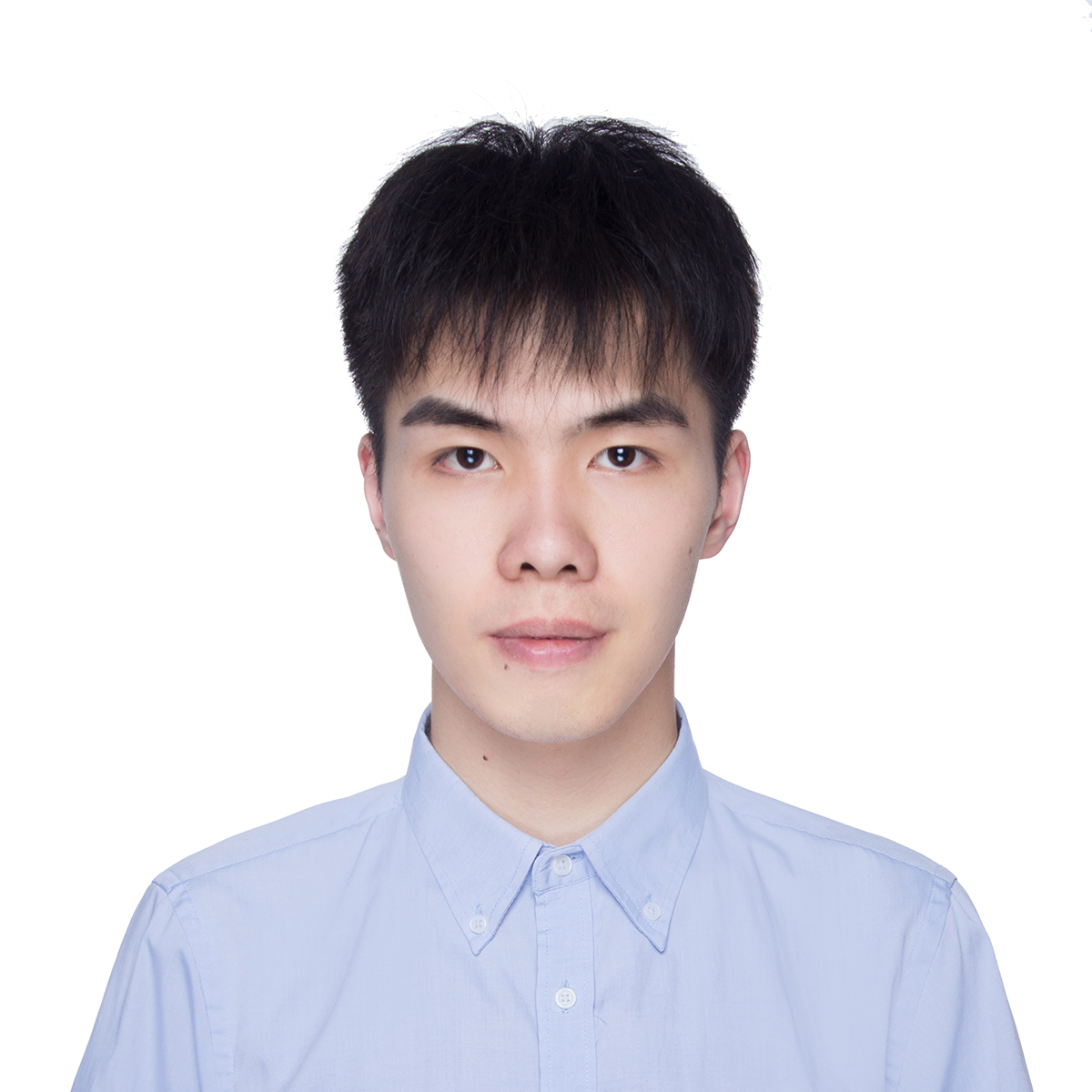}}]{Songqiang Chen} received his bachelor's degree in computer science and technology and his master's degree in software engineering from Wuhan University in 2020 and 2023, respectively. He is currently a Ph.D. student at The Hong Kong University of Science and Technology. 
His research interests lie in software engineering (SE), with a particular focus on AI techniques for software quality enhancement activities such as testing and code analysis. He is also interested in using SE techniques like Metamorphic Testing to validate and improve AI products.
\end{IEEEbiography}

\end{document}